\documentclass[apj]{emulateapj}
\usepackage[colorlinks,linkcolor={blue},citecolor={blue},urlcolor={red}]{hyperref}
\usepackage{epsfig,graphicx,natbib,amsmath,amsfonts,amssymb}%,xfrac}
\usepackage{color}
\usepackage{multirow}
\usepackage{booktabs}
\usepackage{amssymb}
\usepackage{threeparttablex} % for "ThreePartTable" environment
\usepackage{booktabs} 

\usepackage{longtable}
\usepackage{array} % for extrarowheight
\usepackage[dvipsnames]{xcolor}  

\newcommand{\fq}{$f_{\rm Q}$}

\newcommand{\msolar}{${\rm M}_\odot$}
\newcommand{\mstar}{$M_\ast$}
\newcommand{\lgmstar}{$\log_{10}$($M_\ast/h^{-2}$\msolar)}

\newcommand{\dindex}{D$_n$(4000)}

\newcommand{\Msun}{{\rm  M}_{\odot}}
\newcommand{\sersic}{S\'{e}rsic}

\newcommand{\myemail}{\email{ecwang16@ustc.edu.cn(EW), whywang@ustc.edu.cn(HW)}}

\shorttitle{The dearth of difference between central and satellite galaxies}
\shortauthors{Wang et al.}

\graphicspath{{fig/}}
\begin{document}

\title{The dearth of difference between central and satellite galaxies I. Perspectives on star formation quenching and AGN activities} 
\author{
Enci Wang\altaffilmark{1,2},
Huiyuan Wang\altaffilmark{1,2},
Houjun Mo\altaffilmark{3,4}, 
S.H. Lim\altaffilmark{4},
Frank C. van den Bosch\altaffilmark{5},
Xu Kong\altaffilmark{1,2},
Lixin Wang\altaffilmark{3}, 
Xiaohu Yang\altaffilmark{6,7},
Sihan Chen\altaffilmark{1}
} \myemail

\altaffiltext{1}{CAS Key Laboratory for Research in Galaxies and Cosmology, Department of Astronomy, University of Science and Technology of China, Hefei 230026, China}
\altaffiltext{2}{School of Astronomy and Space Science, University of Science and Technology of China, Hefei 230026, China}
\altaffiltext{3}{Tsinghua Center of Astrophysics \& Department of Physics, Tsinghua University, Beijing 100084, China}
\altaffiltext{4}{Department of Astronomy, University of Massachusetts, Amherst MA 01003-9305, USA}

\altaffiltext{5}{Department of Astronomy, Yale University, P.O. Box 208101, New Haven, CT 06520-8101, USA}
\altaffiltext{6}{Department of Astronomy, Shanghai Jiao Tong University, Shanghai 200240, China}
\altaffiltext{7}{IFSA Collaborative Innovation Center, Shanghai Jiao Tong University, Shanghai 200240, China}
%\altaffiltext{4}{Key Laboratory for Research in Galaxies and Cosmology, Shanghai Astronomical Observatory, 
%Chinese Astronomical Society, 80 Nandan Road, Shanghai 200030, China}

\begin{abstract}  
  We investigate the quenching properties of central and satellite
  galaxies, utilizing the halo masses and central-satellite
  identifications from the SDSS galaxy group catalog of Yang et al. We
  find that the quenched fractions of centrals and satellites of
  similar stellar masses have similar dependence on host halo
  mass. The similarity of the two populations is also found in terms
  of specific star formation rate and 4000 \AA\ break.  The quenched
  fractions of centrals and satellites of similar masses show similar
  dependencies on bulge-to-total light ratio, central velocity
  dispersion and halo-centric distance in halos of given halo
  masses. The prevalence of optical/radio-loud AGNs is found to be
  similar for centrals and satellites at given stellar masses.  All
  these findings strongly suggest that centrals and satellites of
  similar masses experience similar quenching processes in their host halos. We discuss implications of our results for the understanding of galaxy quenching.
  
\end{abstract}

\keywords{galaxies: general -- methods: observational}

\section{Introduction}
\label{sec:introduction}
 
Large imaging and spectroscopic surveys of galaxies have revealed
a pronounced bimodality in the distributions of the rest-frame color
and star formation rate of galaxies \citep[e.g.][]{Strateva-01, Baldry-04,
  Bell-04, Blanton-05b, Faber-07, Wetzel-Tinker-Conroy-12}.
%Bimodal distributions have been found in the rest-frame color and star
%formation rate of galaxies \citep[e.g.][]{Strateva-01, Baldry-04,
%  Bell-04, Blanton-05b, Faber-07}, from large imaging and
%spectroscopic surveys of galaxies \citep[e.g.][]{York-00,
%  Colless-01}.
Hence, galaxies are naturally divided into a population of star
  forming galaxies, and a quenched (or quiescent) population.
Galaxies in the star-forming population typically have on-going star
formation activity with disk-like morphology, while the
quenched population typically reveals little to no ongoing star formation, and a spheroid-like morphology \citep{Strateva-01,
  Kauffmann-03, Baldry-04, Brinchmann-04, Li-06, Muzzin-13,
  Barro-17}. This bimodality persists out to redshift of at
least 2.5 \citep[e.g.][]{Bundy-06, Faber-07, Martin-07, Brammer-09,
  Muzzin-12, Huang-13}. In addition the prevalence of the quenched
fraction has increased significantly since redshift $z\sim 1$
\citep{Muzzin-13, Tomczak-14, Barro-17}, indicating that star
formation cessation is one of the theme songs in galaxy evolution over
the past 8 Gyr.  Following convention in the literature, we refer to
this star formation cessation as ``quenching''.
    
Although a complete picture of star formation quenching is still
  lacking, recent studies have shown that the quenched fraction of
galaxies exhibits strong dependence both on the galaxy's
  internal structure, such as bulge-to-total ratio, \sersic\ index,
central stellar mass density and central velocity dispersion
\citep[e.g.][]{Driver-06, Cameron-09, Wuyts-11, Mendel-13, Fang-13,
  Bluck-14, Lang-14, Bluck-16, Barro-17, Wang-17, Wang-18b}, and on properties
of the environment within which the galaxy resides, such as
local matter/galaxy density, morphology of the large scale structure,
and host halo mass \citep[e.g.][]{Balogh-04, Weinmann-06,
  vandenBosch-08, Peng-10, Peng-12, Wetzel-13, Woo-13, Bluck-14,
  Woo-15, Wang-16}.  These findings suggest that both internal
and environmental mechanisms may be important in quenching star
formation.
    
In the current cold dark matter (CDM) cosmogony, galaxies are assumed
to form and evolve within dark matter halos
\citep[e.g.][]{White-Rees-78,Mo-vandenBosch-White-10}. In a halo, the
dominant galaxy, which usually resides near the halo center
\citep[e.g.,][]{Lange-18}, is referred to as the central galaxy, while
other galaxies that orbit the central are referred to as
satellites. Central and satellite galaxies are often investigated
separately, as the dominating quenching mechanisms for the two
populations are believed to be different
\citep[e.g.][]{Weinmann-06, vandenBosch-08, Peng-10, Woo-13, Bluck-14, Woo-15, Knobel-15}.
   
For centrals, the quenched fraction is found to be strongly correlated 
with their internal structure, but only weakly with the environmental 
density and host halo mass \citep{Peng-10,Fang-13, Bluck-14, Woo-15}. 
For example, the presence of a massive bulge appears to be a 
necessary condition for quenching star formation, as claimed in  
\citet{Bell-08}, \citet{Cheung-12} and \citet{Fang-13}.
%   By investigating the relation between quenched fraction and structure properties for over half million SDSS galaxies,  \cite{Bluck-14} further found the bulge mass is a key factor in quenching star formation rather than bulge-to-total stellar mass ratio.  More recently, by applying an artificial neural network (ANN) approach to over 400,000 central galaxies,
More recently, \cite{Teimoorinia-Bluck-Ellison-16} argued that
the central velocity dispersion of galaxies is more closely linked to
the cessation of star formation than any other variable
considered, including bulge mass and halo mass, suggesting that
quenching is related to the central supermassive black holes through
active galactic nucleus (AGN) feedback \citep[e.g][]{Croton-06,
  Henriques-15, Schaye-15}.  Indeed, hydrodynamic simulations have
shown that the AGN-driven radiation, jets and winds can lead to the
ejection or heating of the interstellar medium of galaxies and/or
intra-cluster medium, thereby suppressing star formation and
maintaining quiescence \citep[e.g.][]{Sijacki-07, Dubois-12,
  Cicone-14}. However, observational evidence for star formation
quenching by AGN feedback remains elusive. In fact, both
positive and negative feedback effects have been claimed in the
  literature \citep[e.g.][]{Fabian-12, Mullaney-12, Mullaney-15,
  Delvecchio-15, Rodighiero-15, Mahoro-Povic-Nkundabakura-17,
  Kalfountzou-17}.  Finally, in halos with massed above a few
$10^{12}\Msun$, shock heating may effectively reduce gas cooling,
thereby reducing star formation efficiency in central galaxies
\citep{Rees-Ostriker-77, Birnboim-Dekel-03, Dekel-Birnboim-06},
although it is not obvious whether this can explain the
dependence of quenching on the intrinsic properties of galaxies.
    
For satellites, a number of additional `satellite-specific'
  quenching processes have been suggested, including galaxy mergers
\citep[e.g.][]{Conselice-Chapman-Windhorst-03, Cox-06, Cheung-12},
tidal interaction \citep{Toomre-Toomre-72, Read-06}, ram-pressure
stripping \citep[e.g.][]{Gunn-Gott-72, Abadi-Moore-Bower-99,
  Hester-06, Wang-15}, strangulation \citep{Larson-80,
  Balogh-Navarro-Morris-00, vandenBosch-08, Weinmann-09}, and galaxy
harassment \citep{Farouki-81, Moore-96, Moore-Lake-Katz-98}.
Different from centrals, the quenched fraction of satellites has been found to depend on the number density of surrounding galaxies
\citep{Peng-10, Peng-12}. More recently, \cite{Woo-15} found that
quenching of satellites is strongly correlated with the distance to
cluster/group center, indicating that denser environments are more
effective at quenching galaxies \citep[see also][]{Gomez-03,
  Balogh-04, Blanton-Roweis-07, Haines-07, Wolf-09,
  Wetzel-Tinker-Conroy-12, Woo-13}. Note, though, that such a
  trend may also arise from the stellar mass dependence of quenching
  combined with mass segregation \citep[][]{vandenBosch-08b}, and/or
  from the fact that satellites at smaller halo-centric distances
  have, on average, been accreted earlier
  \citep[e.g.,][]{vandenBosch-16}, and therefore exposed to
  satellite-specific quenching processes longer.

Given that environmental effects may affect centrals and satellites
differently, a comparison between the two galaxy populations in their
star formation properties may provide an avenue to understand the
importance of environmental quenching \citep{vandenBosch-08,
  Wetzel-Tinker-Conroy-12, Peng-12, Hirschmann-14,
  Knobel-15,Spindler-Wake-17, Wang-18}.  Indeed, numerous earlier
investigations have found that satellites tend to be more quenched
than centrals of the same stellar mass \citep{vandenBosch-08,
  Wetzel-Tinker-Conroy-12, Peng-12, Woo-13, Bluck-14, Bluck-16}.
These results have been widely interpreted as evidence for
  satellites experiencing some specific quenching processes.
However, there are also indications that centrals and
satellites may actually not be that different.  For example,
\cite{Hirschmann-14} found that centrals and satellites show similar
quenching behavior with local densities, when dividing galaxies into a
series of narrow stellar mass bins. \cite{Knobel-15} found that
centrals that have massive satellites respond to environments in the
same way as satellites of the same stellar mass. More recently,
\cite{Wang-18} analyzed the environmental quenching efficiency that
quantifies the quenched fraction as a function of halo mass, and found
that centrals and satellites respond to their halo masses in a similar
way. Furthermore, they found that the difference between centrals and
satellites seen in previous investigations arises largely from the
fact that centrals and satellites of the same stellar mass reside, on
average, in halos of different masses. These results strongly
  suggest that host halo mass is the prime environmental parameter
  that regulates the quenching of both centrals and satellites.
% These results strongly suggest that centrals and satellites have
% experienced similar quenching processes in their host halos.

As many galaxy properties are correlated, well controlled samples are
needed in order to investigate whether or not the differences seen
between centrals and satellites are truly due to different
evolutionary processes instead of due to sample selections. In this
paper, we extend the analysis of our previous work \citep{Wang-18} and
present a comprehensive comparison between the central and satellite
populations in a number of properties related to star formation and
quenching. These include the quenched fraction, the 4000 \AA\ break,
specific star formation rate and the prevalence of optical/radio-loud
AGNs for galaxies controlled both in stellar mass and host halo mass,
as well as the quenched fraction as a function of bulge-to-total
ratio, galaxy central velocity dispersion and halo-centric radius.
The remainder of the paper is structured as follows.  Section
\ref{sec:data} presents the observational data and the definitions of
physical properties we use in our analysis.  We present our results
for star formation quenching in \S\ref{sec:results}, and for AGN
activities in \S\ref{sec:AGN}. Finally, we summarize our results and
discuss their implications in \S\ref{sec:summary}. The cosmology used
is that of WMAP3 \citep{Spergel-07}: $\Omega_{m}=0.238$,
$\Omega_{\Lambda}=0.762$ and $h=0.73$, which is the same as that
adopted for the group catalog used here \citep{Yang-07}.

\section{Observational data}
\label{sec:data}

\subsection{Galaxies}
\label{subsec:catalog}
\subsubsection{The galaxy sample}
   
Our galaxy sample is selected from the New York University Value Added
Galaxy Catalog\footnote{http://cosmo.nyu.edu/blanton/vagc}
\citep[NYU-VAGC;][]{Blanton-05a} of the SDSS DR7 \citep{Abazajian-09}.
We selected galaxies with redshift in the range of $0.01<z<0.2$, with
spectroscopic completeness ($C$) greater than 0.7, and with the
$r$-band flux-limited of $r=17.72\,{\rm mag}$, which result in 544,328
galaxies.  The first two criteria ensure that the selected galaxies
are the same as those used in the construction of the group catalog
\citep{Yang-07, Yang-Mo-vandenBosch-09} adopted here.  We combine the
sample with MPA-JHU
catalog\footnote{http://wwwmpa.mpa-garching.mpg.de/SDSS/DR7}
\citep{Kauffmann-03,Brinchmann-04} and the bulge+disk decomposition
catalog of \cite{Simard-11} to obtain derived quantities of individual
galaxies, with excluding the unmatched sources ($\sim3.6\%$).

\begin{figure*}
\epsfig{figure=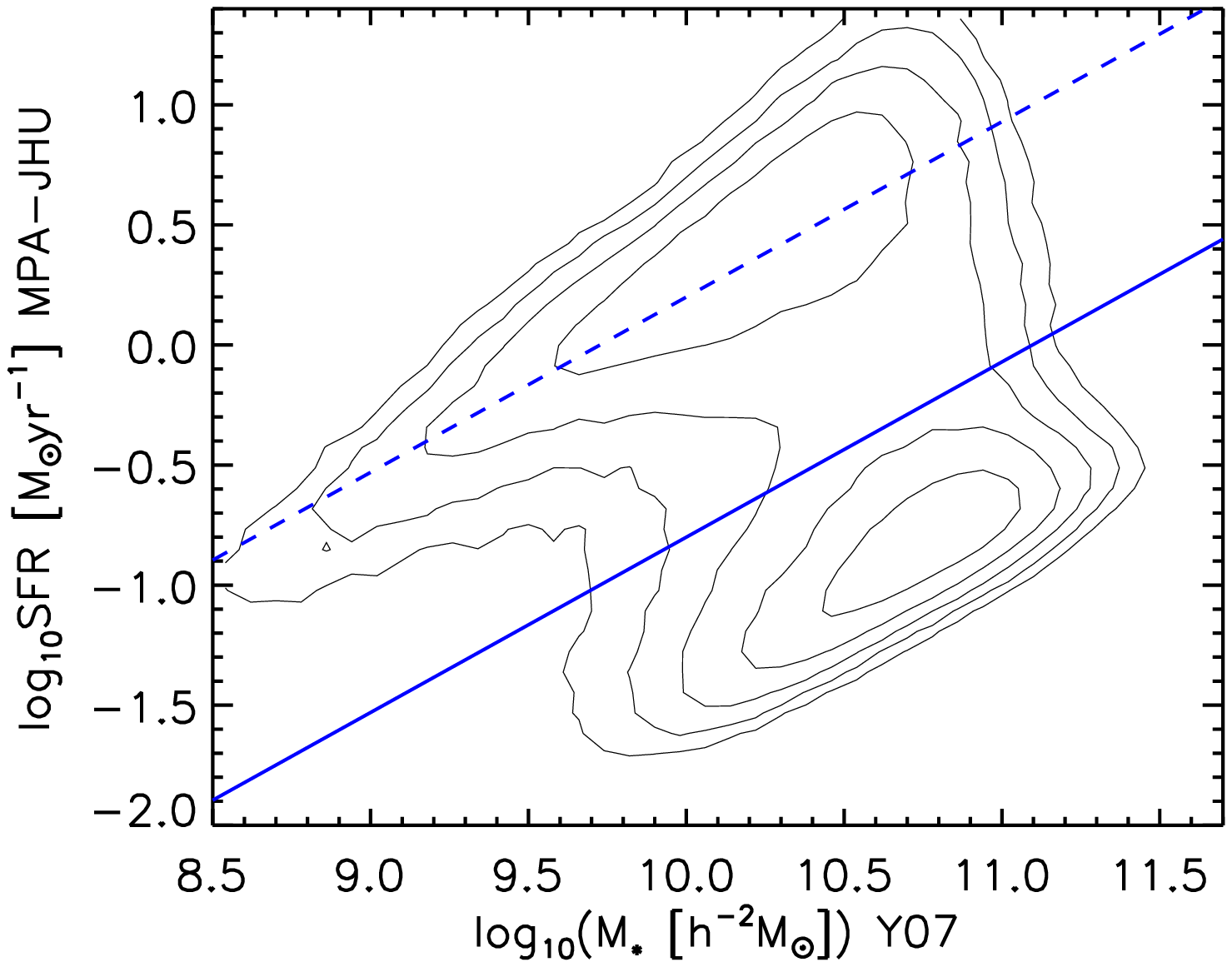,clip=true,width=0.33\textwidth}
\epsfig{figure=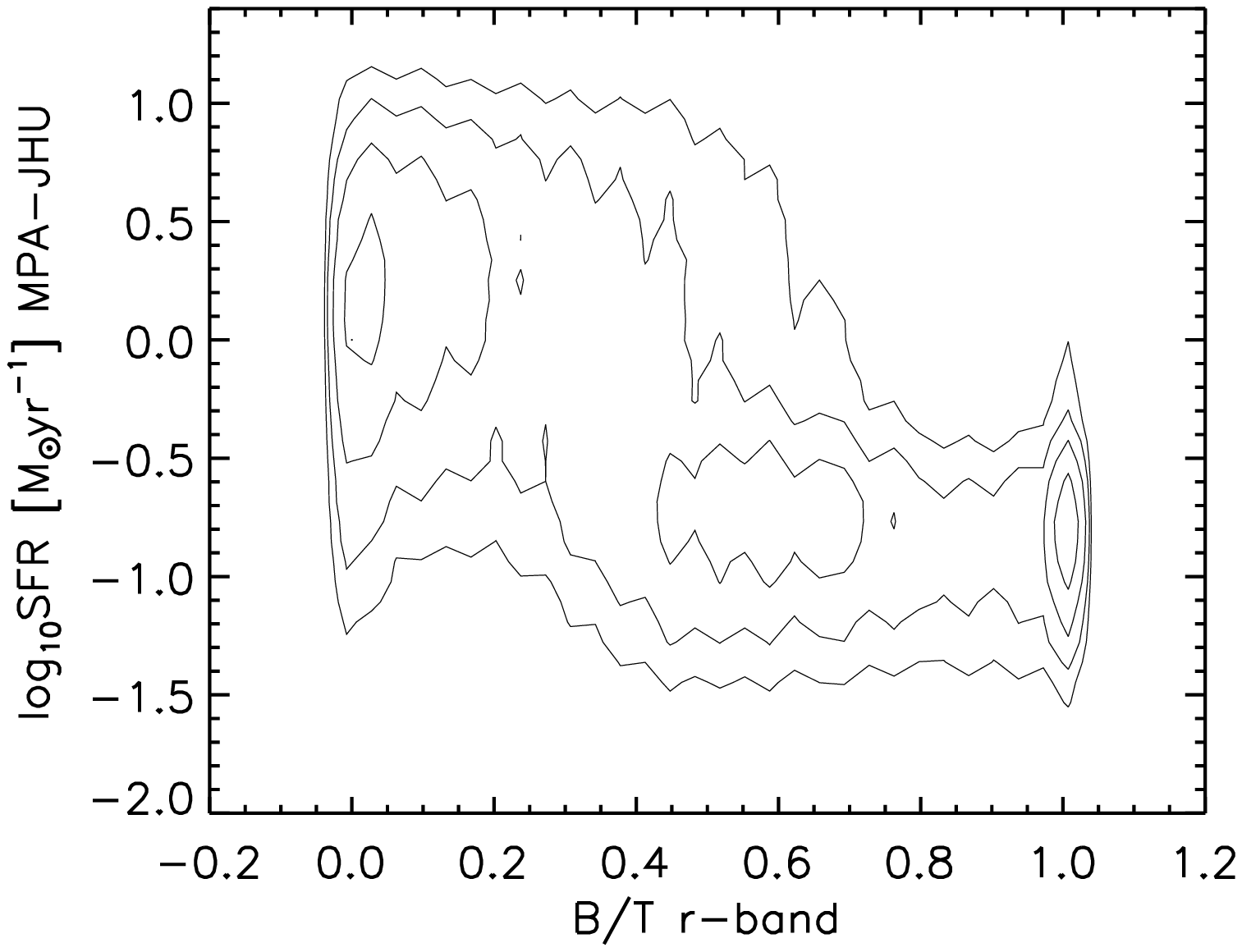,clip=true,width=0.33\textwidth}
\epsfig{figure=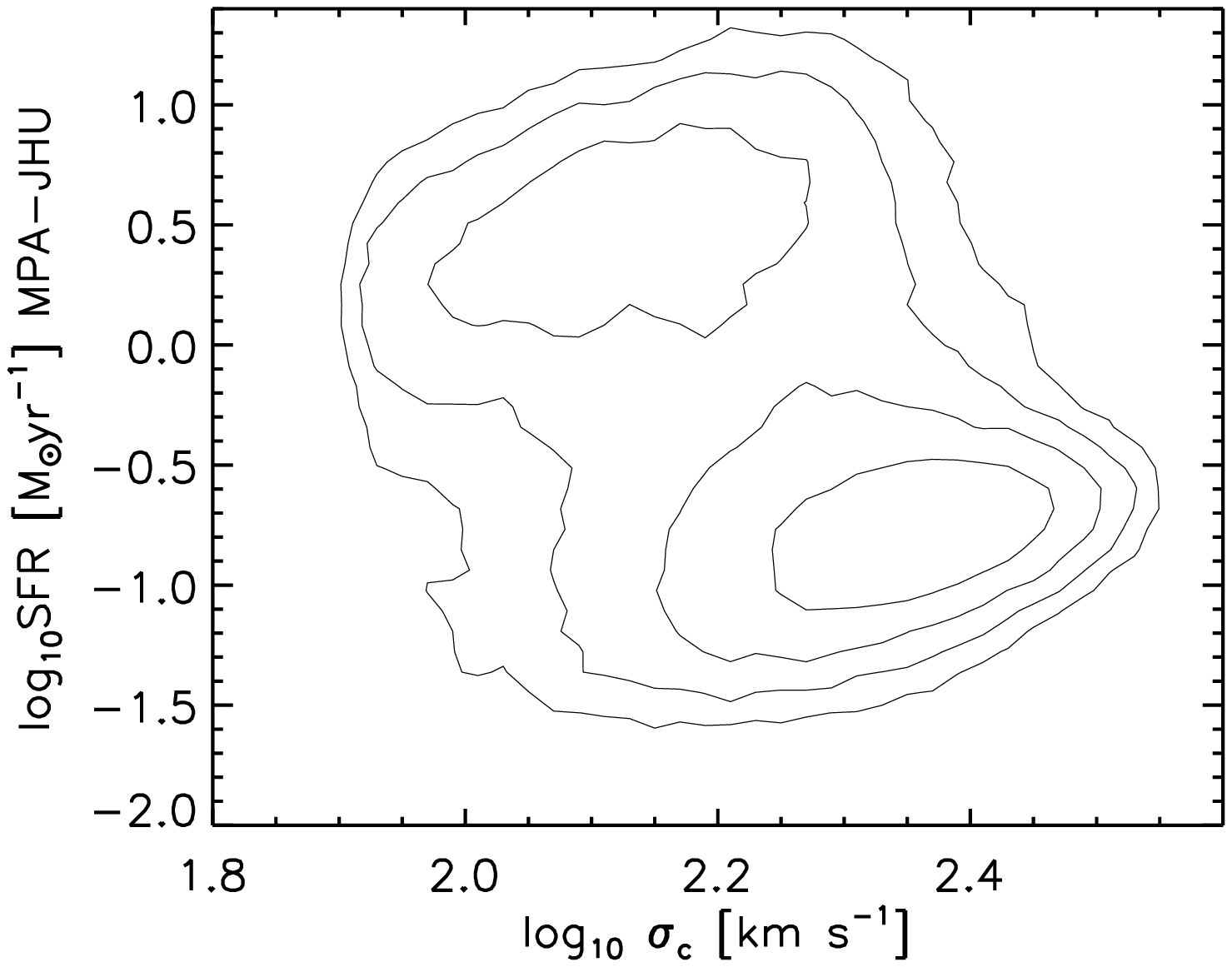,clip=true,width=0.33\textwidth}
\epsfig{figure=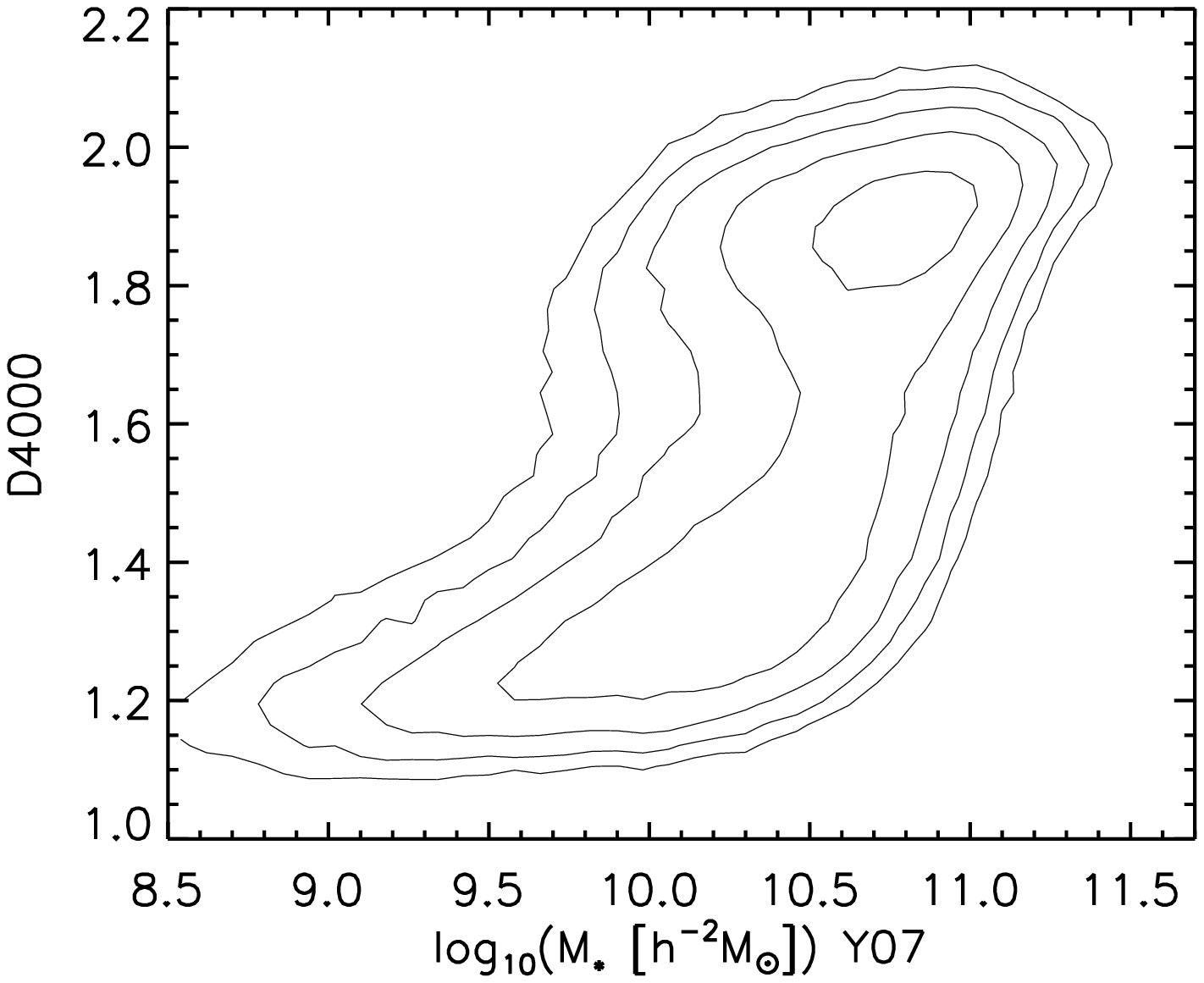,clip=true,width=0.33\textwidth}
\epsfig{figure=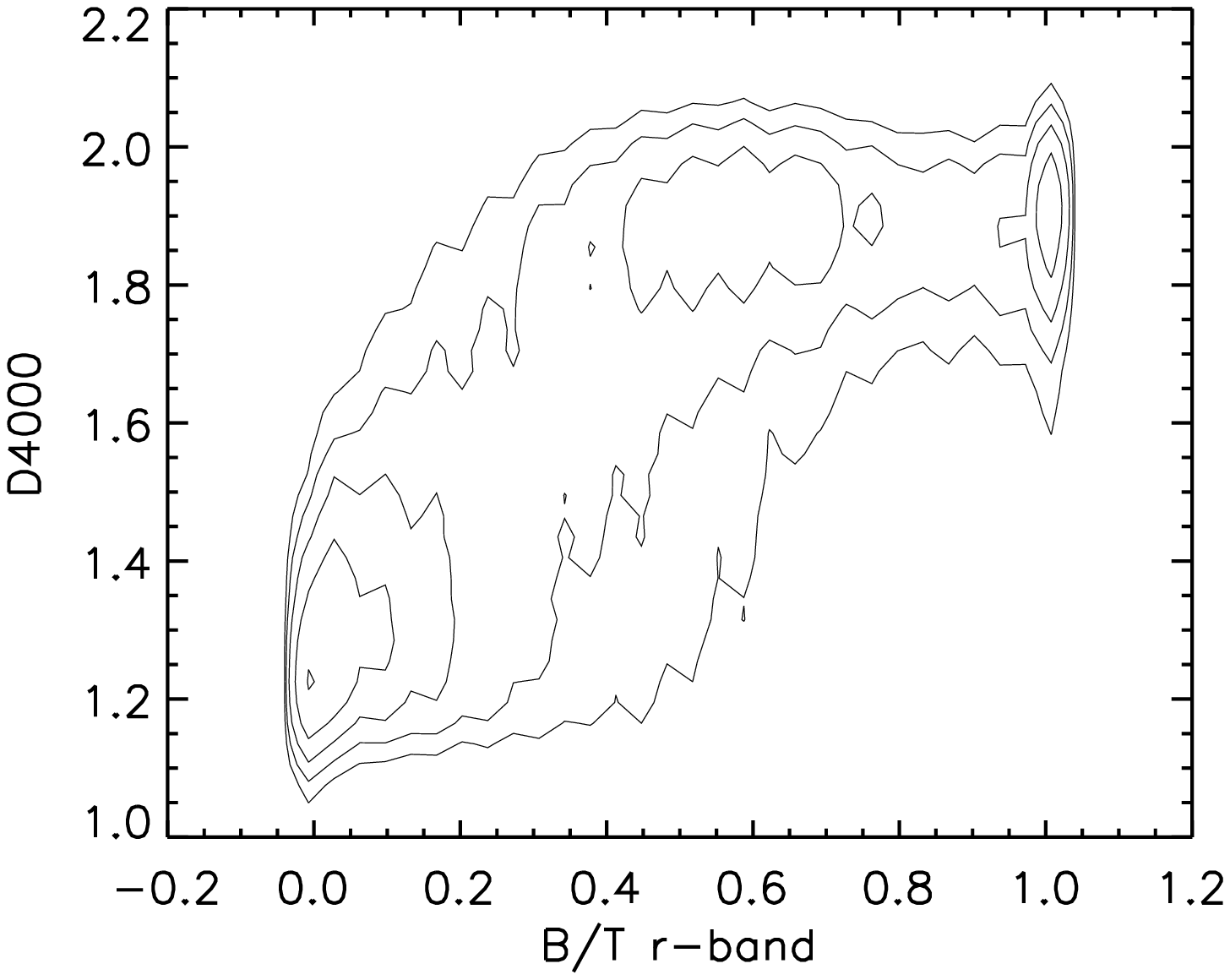,clip=true,width=0.33\textwidth}
\epsfig{figure=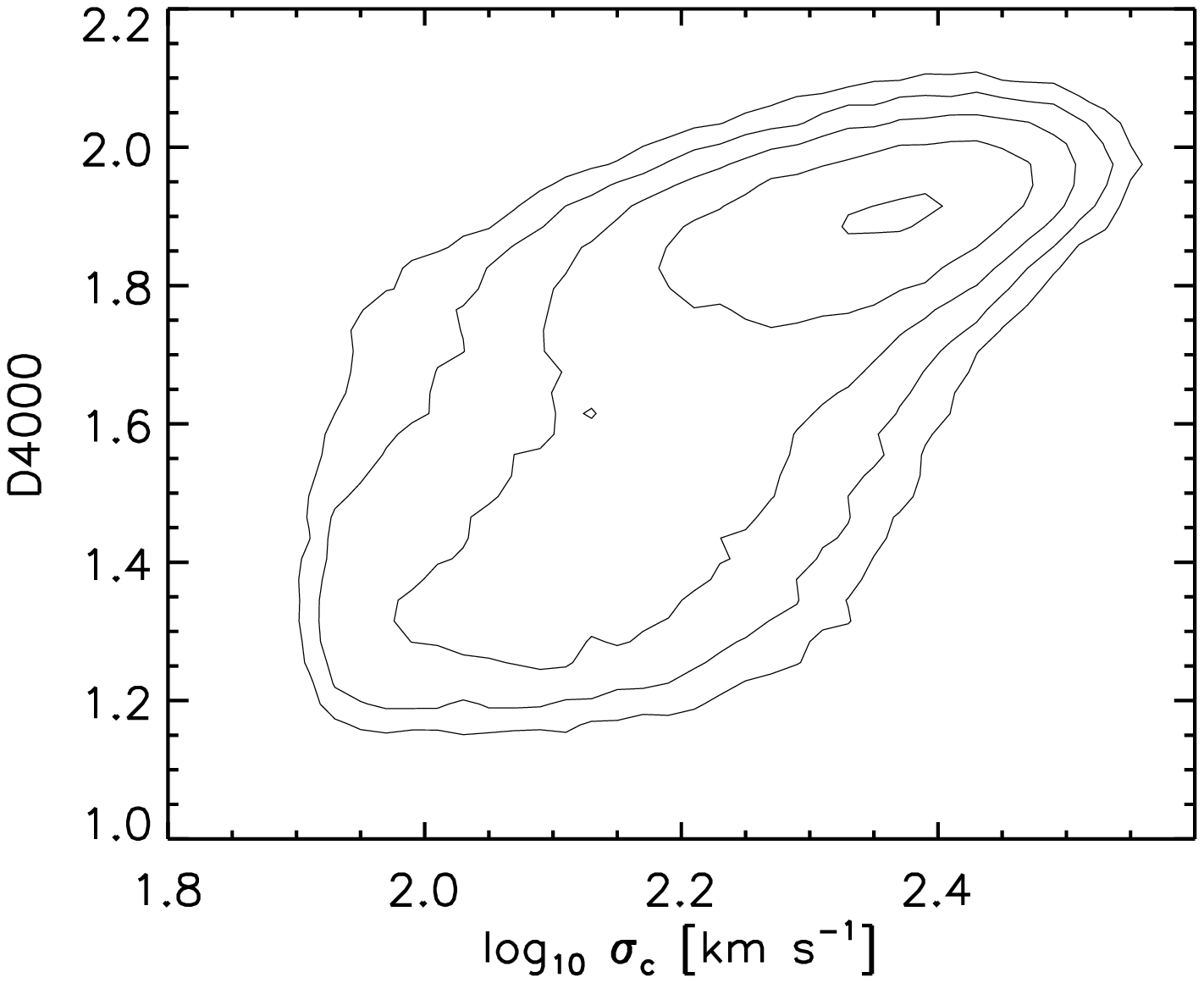,clip=true,width=0.33\textwidth}
\caption{Star formation rate (top panels) and the 4000\AA\ break (bottom panels) as a function of stellar mass, bulge-to-total light ratio and central velocity dispersion for individual galaxies in the whole sample. 
In the top left panel, the blue dashed line represents the star formation main sequence taken from \cite{Bluck-16}, and the blue solid line, which is 1 dex lower than the former, is used to separate the star-forming and the quenched populations (see details in \S\ref{subsec:qf}).}
\label{fig:quench_def}
\end{figure*}

The MPA-JHU catalog provides the measurements of the main physical parameters 
used in this paper, such as star formation rate (SFR), 4000 \AA\ break 
(\dindex), stellar velocity dispersion ($\sigma_*$), and the emission line 
flux. The SFRs are measured by an updated version of the method of 
\cite{Brinchmann-04} using the Kroupa initial mass 
function \citep{Kroupa-Weidner-03}. The 4000 \AA\ break is defined as the 
ratio of the flux between the red and blue continua at 4000 \AA\ 
\citep{Balogh-99}. The measurements of the stellar velocity dispersion 
are based on the SDSS 3-arcsec fiber spectra of the galaxy center. 
We correct the measurements to the same effective aperture, using the 
formula of \cite{Cappellari-06}:
\begin{equation}
\sigma_{\rm c}=\left(\frac{R_{\rm e}/8}{R_{\rm ap}}\right)^{-0.066}\sigma_{\rm ap},
\end{equation} 
where $R_{\rm ap}$ is the aperture radius, $\sigma_{\rm ap}$ the
velocity dispersion measured within the aperture, and $R_{\rm e}$ the
effective radius taken from the NYU-VAGC. The factor of $1/8$ is
chosen to be consistent with the measurements in the literature.  The
catalog of \cite{Simard-11} provides the bulge-to-total light ratios
of individual galaxies obtained from the decomposition of the SDSS
$r$-band imaging data, with the assumption of a \sersic\ ($n_{\rm
  b}=4$) bulge plus an exponential disk. 
Figure \ref{fig:quench_def} presents the SFR and \dindex\ as a function of stellar mass, bulge-to-total light ratio and central velocity dispersion, which provides a global impression of the distributions of these parameters for the sample galaxies. 

Our final sample contains 524,852 galaxies with measured SFRs,
among which about $\sim 24\%$ are satellite galaxies. Since the sample is flux-limited at $r=17.72\,{\rm mag}$, we assign each galaxy a weight $w=(V_{\rm max}C)^{-1}$ to correct for Malmquist bias and redshift incompleteness. We use these weights throughout, when deriving statistics for our sample of galaxies. Here $V_{\rm max}$ is the comoving volume between the minimum redshift, $z_{\rm min}=0.01$, and the maximum redshift, $z_{\rm max}$, out to which the galaxy falls within the flux limit of the survey, and is computed using the $K$-correction utilities (v4\_2) of \cite{Blanton-Roweis-07}.

\subsubsection{AGN properties}

The optical AGNs used here are identified on the basis of the BPT
diagram \citep{Baldwin-Phillips-Terlevich-81}, which can be used to
separate type II AGNs (ones with only narrow lines) from star forming
galaxies and composite galaxies \citep{Kewley-01}.  The fluxes of the
relevant emission lines, such as H$\alpha$, H$\beta$,
[OIII]$\lambda$5007 and [NII]$\lambda$6583, are taken from the MPA-JHU
catalog \citep{Brinchmann-04}.  We only identify strong AGNs (Seyfert
galaxies) using the Seyfert-LINER\footnote{Low Ionization Nuclear
  Emission-line Region} demarcation given by \citet{CidFernandes-10}.
Following \cite{Pasquali-09}, we require the relevant emission lines
to have signal-to-noise ratios greater than $3.0$, and the emission
line fluxes are corrected for intrinsic extinction based on the Balmer
decrement and a dust attenuation curve of the form $\lambda^{-0.7}$
\citep{Charlot-Fall-00}, assuming an intrinsic ratio ${\rm
  H}\alpha/{\rm H}\beta=2.86$ \citep{Osterbrock-89}.

The radio-loud AGNs are obtained by matching the galaxies in our
sample with the radio galaxy catalog of \cite{Best-Heckman-12}. This radio galaxy catalog is constructed by cross-matching the MPA-JHU catalog with both the NVSS \citep[the National Radio Astronomy Observatory (NRAO) Very Large Array (VLA) Sky Survey;][]{Condon-98} and FIRST \citep[the Faint Images of the Radio Sky at Twenty centimetres;][]{Becker-White-Helfand-95} survey following the method of \cite{Best-05}. The radio catalog has a flux-density limit of 5 mJy, which corresponds to a luminosity of $L_{\rm 1.4 GHz} \sim 10^{23}$ W Hz$^{-1}$ at $z=0.1$.
We therefore restrict our sample to $z<0.1$ and cross-match with
sources with $L_{\rm 1.4 GHz} > 10^{23}$ W Hz$^{-1}$.  Radio sources
associated with star-forming galaxies are not included in the sample, as the classification used in \cite{Best-Heckman-12} puts radio-AGNs and star-forming galaxies into two mutually exclusive classes. The $V_{\rm max}$ correction is recalculated for 
the resulting sample, which is used only for calculating the 
radio loud-AGN fractions of centrals and satellites. 

\subsection{Groups of galaxies}
\label{subsec:group}

The galaxy groups used in our analysis are taken from the SDSS
 DR7 group catalog of \citet{Yang-07}, which is based on the
 halo-based group finding algorithm developed in \citet{Yang-05}. For each galaxy in our sample, this group catalog provides properties of the inferred host halo (e.g., mass and size), and indicates whether the galaxy is a central or a satellite.  The WMAP3 cosmology \citep{Spergel-07} is assumed both in the group finder and in calculating quantities of groups and member galaxies. For each galaxy group, two different halo masses are assigned; one based on its characteristic stellar mass and one based on its characteristic luminosity.  
Here we use the former definition, and identify the central galaxy to be the most massive one in a given group, as recommended in the original papers. For $\sim 22\%$ of all groups, no halo mass is available due to limitations of the ranking-based halo mass assignment method. Since these groups are the least massive ones, we assign them to our lowest halo mass bin ($M_{\rm h}<10^{12}h^{-1}$\msolar) in the statistics that follow.
Finally, in order to reduce boundary effects, we exclude groups with $0<f_{\rm edge}<0.7$, where $f_{\rm edge}$ is the fraction of the volume of a group that lies within the survey boundary. We refer the reader to \cite{Yang-05} and \cite{Yang-07} for details.

The stellar masses used for the group finder were taken from
the NYU-VAGC \citep{Blanton-05a}, and were calculated using the
relation between the stellar mass-to-light ratio and the $g-r$ color,
as given in \cite{Bell-03}. Since the halo masses are based on these
stellar mass estimates, we use these stellar masses for all our
analyses. The halo radius of a group is estimated as
\begin{equation}
r_{\rm 180}=1.26 h^{-1} {\rm Mpc} \left(\frac{M_h}{\rm 
10^{14}h^{-1}M_{\odot}}\right)^{1/3}(1+z_{\rm group})^{-1}, 
\end{equation}
where $z_{\rm group}$ is the redshift of the group center, and $r_{\rm
  180}$ is the radius within which the dark matter halo has an
overdensity of $180$. For each galaxy, we define a scaled halo-centric
radius $R_{\rm p}/r_{\rm 180}$, which is the projected distance from
the galaxy to the host group center in units of the halo virial radius
of the host group. The group center used here is the
luminosity-weighted center of galaxies, which means that central
galaxies do not always locate in the group center.

\subsection{Statistical quantities} 
\label{subsec:qf}

The bimodal distribution of galaxies in the color-magnitude diagram is 
observed both in the local universe and in high redshift 
\citep[e.g.][]{Strateva-01, Baldry-04, Faber-07, Erfanianfar-16}. Based on this 
bimodality, galaxies can be divided into a star-forming (SF) population 
and a quenched population according to the color-magnitude diagram 
or the SFR-stellar mass diagram.  
As shown in Figure \ref{fig:quench_def},  a bimodal distribution is seen in each of the panels, indicating how the star-forming and quenched populations are separated in the corresponding parameter space.
The top left panel of Figure \ref{fig:quench_def} 
shows the SFR - stellar mass relation for galaxies in our sample. 
To separate galaxies in our sample into a star-forming population and  
a quenched population, we adopt the demarcation line 
suggested by \cite{Bluck-16}, which is parallel to but 1 dex below 
the star formation main sequence and can be written as:
\begin{equation}
\log_{10}{\rm SFR} = 0.73\log_{10}M_*-1.46\log_{10}h-8.3,
\end{equation}
where the reduced Hubble constant $h$ (the Hubble constant in the units 
of $100\,{\rm km\, s^{-1}\,Mpc^{-1}}$) is included here to convert the 
units of the stellar mass from $\Msun$ used in \cite{Bluck-16} 
to $h^{-2}\Msun$ used here. According to this separation, a quenched 
galaxy is defined to be the one that has a star formation rate 
at least a factor of 10 times lower than that of a typical SF galaxy 
of the same stellar mass. We note that our main result is not sensitive to the definition of quenching. Actually, we have examined the results by adopting a flatter division from \cite{Woo-13} with respect to that of \cite{Bluck-16}, and find that the main result still holds.

For a given subsample (S), the quenched fraction (\fq) is defined as:
\begin{equation}\label{eq:qf}
 f_{\rm Q}=\frac{\sum_{i=1}^{S} w_i\times f_{\rm Q,i}}{\sum_{i=1}^{S} w_i},
\end{equation}
where $f_{\rm Q,i}$ represents the quenched status of the $i$th galaxy in the 
subsample ($f_{\rm Q,i}=1$ if the galaxy is quenched, else $f_{\rm Q,i}=0$), 
and $w_{i}$ is the weight of the $i$th galaxy (see \S\ref{subsec:catalog}). 
For a given subsample, the error of the quenched fraction is estimated by 
using 1000 bootstrap samples. Similarly, the AGN fraction is defined as
\begin{equation} \label{eq:agnf}
f_{\rm AGN}=\frac{\sum_{i=1}^{S} w_i\times f_{\rm AGN,i}}{\sum_{i=1}^{S} w_i}, 
\end{equation}
where $f_{\rm AGN,i}=1$ if the $i$th galaxy hosts an AGN, 
and $f_{\rm AGN,i}=0$ otherwise.

\section{Results of quenched fraction}
\label{sec:results}

\subsection{Dependence on halo mass and stellar mass}
\label{subsec:fq_1th}

\begin{figure*}
  \begin{center}
     \epsfig{figure=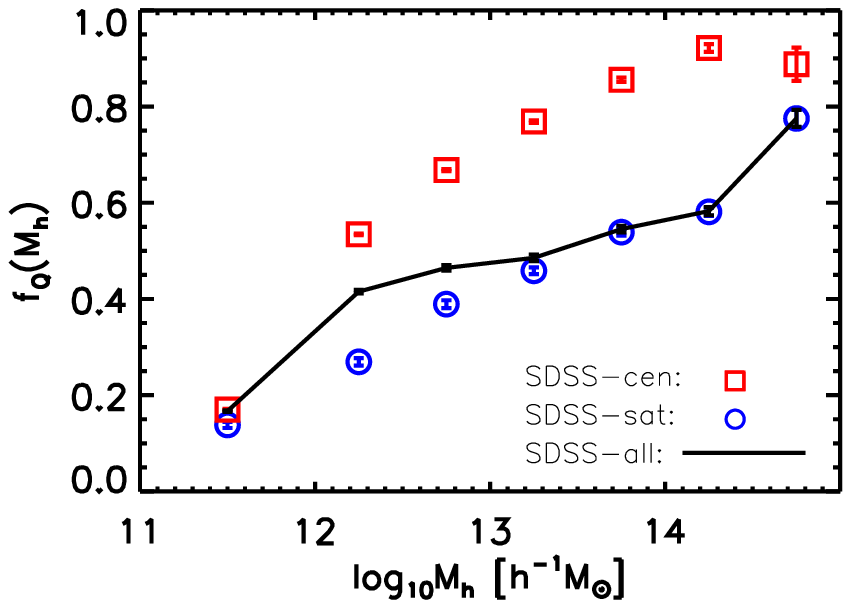,clip=true,width=0.4\textwidth}
     \epsfig{figure=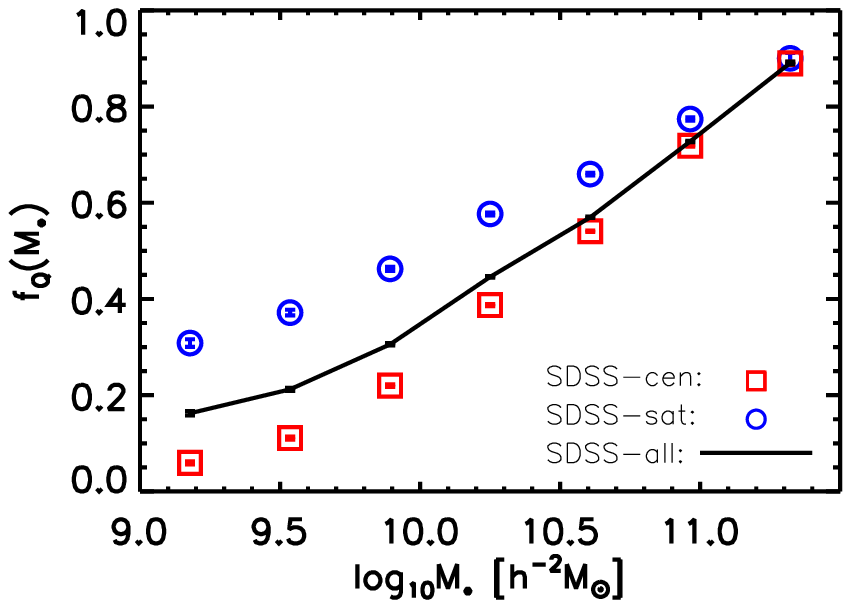,clip=true,width=0.4\textwidth}
   \end{center}
\caption{Left panel: Quenched fraction as a function of halo mass for
  centrals (red squares), satellites (blue circles) and all galaxies
  (black solid line). Right panel: Quenched fraction as a function of
  stellar mass for centrals (red squares), satellites (blue circles)
  and all galaxies (black solid line). The errors are all estimated by
  using the bootstrap method. }
 \label{fig:quench_mh_0th}
\end{figure*}

\begin{figure*}
  \begin{center}
    \epsfig{figure=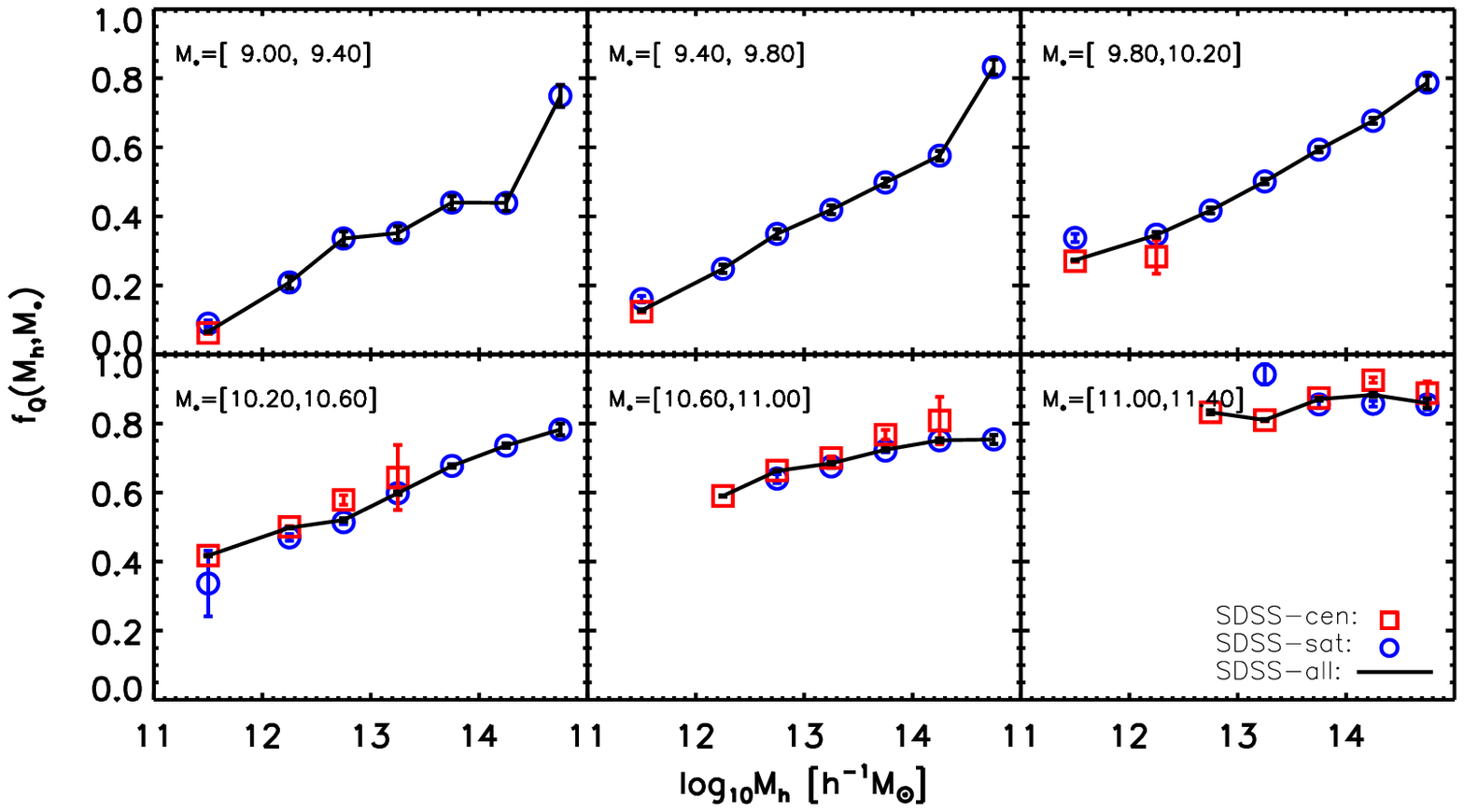,clip=true,width=0.7\textwidth}
    \epsfig{figure=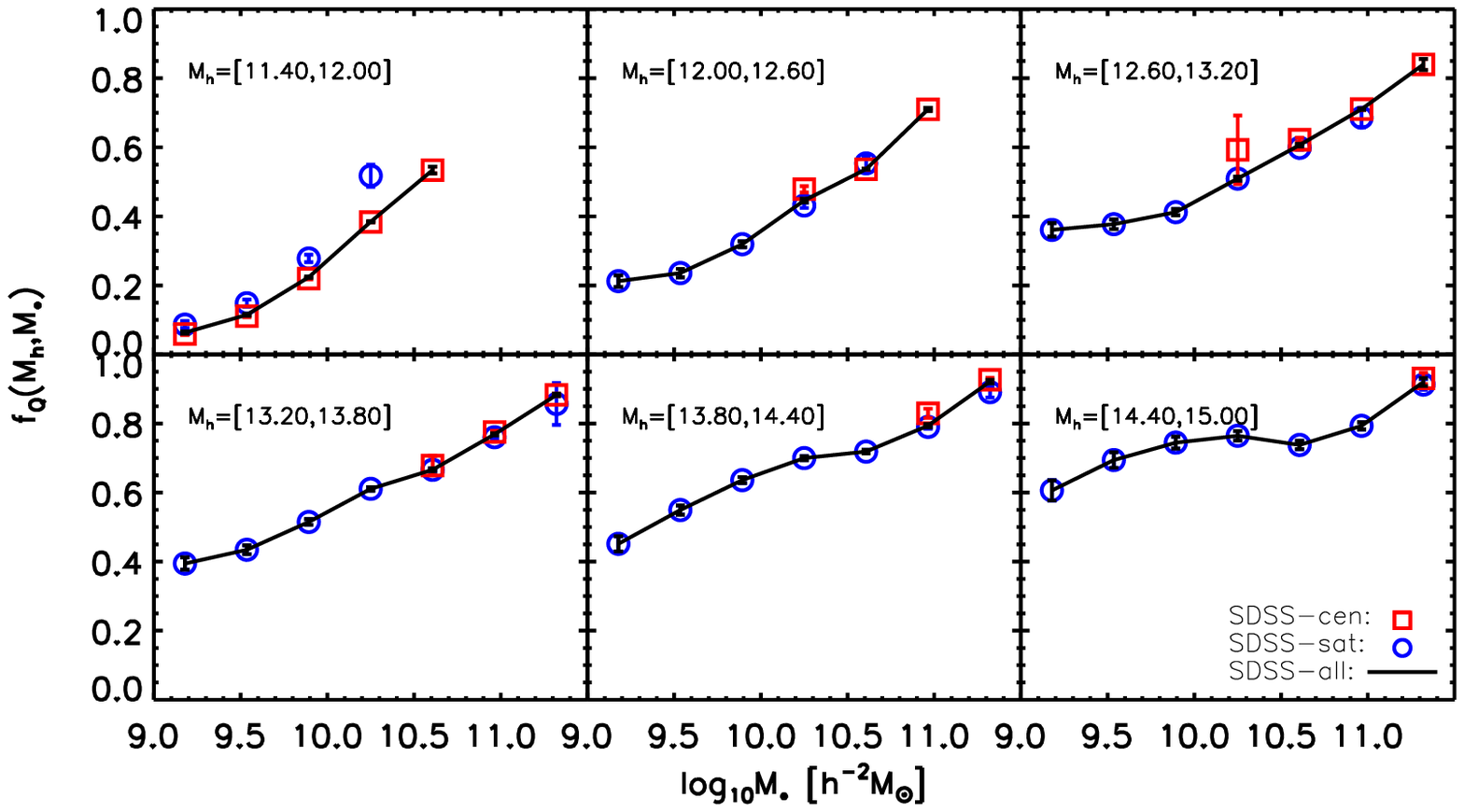,clip=true,width=0.7\textwidth}
    \end{center}
\caption{Top group of panels: Quenched fractions as functions of halo
  mass for centrals (red squares), satellites (blue circles) and all
  galaxies (black lines) in a series of stellar mass bins, as labeled
  in the top-left corner.  Bottom group of panels: Quenched fractions
  as functions of stellar mass for centrals (red squares), satellites
  (blue circles) and all galaxies (black lines) in a series of halo
  mass bins, as labeled in the top-left corner.  }
\label{fig:quench_mh_1th}
\end{figure*}

Stellar mass is one of the most important properties of a galaxy,
which reflect the total amount of stars that formed in the past, while
the host halo mass is one of the most important properties of a dark
matter halo. In the standard model, galaxies are believed to form and
evolve in dark matter halos. Thus, the properties of galaxies are
expected to depend on the properties of their host halos.  Halo mass
has been widely used to link galaxies to dark matter halos through
models such as the halo occupation distribution
\citep[e.g.][]{Jing-Mo-Borner-98, Zheng-05, Li-08}, conditional
luminosity function \citep[e.g.][]{Yang-Mo-vandenBosch-03,
  vandenBosch-07}, abundance matching \citep[e.g.][]{Mo-Mao-White-99,
  Vale-Ostriker-06}, and empirical parameterization
\citep[e.g.][]{Lu-14b, Moster-17}.  More recently, \cite{Wang-18}
showed that halo mass is the primary environmental parameter in
regulating the quenching of centrals and satellites. In this
  section, we follow \cite{Wang-18}, and study the relationships
between the quenched fraction, stellar mass and host halo mass for
centrals and satellites.

Figure \ref{fig:quench_mh_0th} shows the quenched fraction as a
function of halo mass (left panel) and stellar mass (right panel) for
centrals, satellites and all galaxies, respectively. As expected, the
quenched fraction increases with increasing stellar mass and halo mass
for both centrals and satellites, and the result is in good agreement
with previous findings \citep[e.g.][]{Strateva-01, Brinchmann-04,
  Weinmann-06, Wetzel-Tinker-Conroy-12, Woo-13, Bluck-16}.  Moreover,
centrals are more frequently quenched than satellite galaxies at given
halo mass, which is due the fact that centrals are usually more
massive than satellites for a given halo mass.  This trend is reversed
for a given stellar mass \citep[see also][]{vandenBosch-08,
  Weinmann-09, Knobel-13, Bluck-16, Grootes-17, Fossati-17, Wang-18}, which is
usually taken as the evidence for ``environmental quenching'' of
satellite galaxies.  

%However, the comparison shown in Figure \ref{fig:quench_mh_0th} is unfair. Indeed, when comparing centrals and satellites at a given halo mass, we are comparing the two populations with different stellar mass; similarly, when comparison is made at a given stellar mass, we are comparing centrals and satellites in halos of different masses.

However, when comparing centrals and satellites at a given halo
  mass, we are comparing two populations with different stellar mass;
  similarly, when comparison is made at a given stellar mass, we are
  comparing centrals and satellites in halos of different mass. Hence,
  it is not clear whether the differences between centrals and
  satellites arise from `being a satellite' versus `being a central',
  or from the fact that quenching depends on halo mass and/or stellar
  mass. Put differently, the data presented above does not rule out,
  for example, a scenario in which there are no satellite-specific
  processes that cause quenching; rather, quenching is
  (probabilistically) governed by the mass of the host halo, and
  operates equally on centrals and satellites. We can test this,
  though, by comparing the quenched fractions of centrals and
  satellites that are controlled for {\it both} stellar stellar {\it
    and} halo mass.

%Since the quenched fraction may depend on both stellar mass and halo
%mass, it is essential to compare centrals and satellites of similar
%stellar and halo masses in order to examine whether centrals and
%satellites are really different in their quenching properties

To do this, we separate galaxies into six stellar mass bins of the
same width in logarithmic space. The quenched fraction as a function
of halo mass for centrals and satellites are shown in the top group of
panels in Figure \ref{fig:quench_mh_1th}.  The large differences
between centrals and satellites seen in Figure \ref{fig:quench_mh_0th}
are very much reduced here when comparisons are made in narrow stellar
mass bins. Here we see that centrals and satellites have similar
fractions of quenched population at given stellar and halo masses. The
quenched fraction increases with increasing halo mass over the entire
stellar mass range, except the most massive bin.  The dependence of
the quenched fraction on halo mass becomes weaker as the stellar mass
increases, which may indicate a transition of the dominated quenching
mechanism from environmental to internal processes \citep{Peng-10,
  Woo-15}.

We also form samples by dividing galaxies into six uniform halo mass
bins in logarithmic space and present the quenched fractions as a
function of stellar mass in the bottom group of panels in Figure
\ref{fig:quench_mh_1th}. Quenched fraction increases with increasing
stellar mass for both centrals and satellites over the whole stellar
mass range. Centrals and satellites exhibit virtually identical
  trends in the \fq-$M_\ast$ relation in all bins of halo mass.  The
dependence of the quenched fraction on stellar mass become weaker as
halo mass increases. 

Using a larger galaxy sample, we have thus confirmed the result of
\cite{Wang-18}, that centrals and satellites at a given stellar mass
have similar \fq-$M_{\rm h}$ relations. Although this can be naturally 
interpreted as evidence that central and satellite
galaxies in a given halo are quenched by similar physical processes,
we note that there are several alternative interpretations, which are discussed in more detail in \S\ref{sec:summary}. 

\begin{figure*}
  \begin{center}
    \epsfig{figure=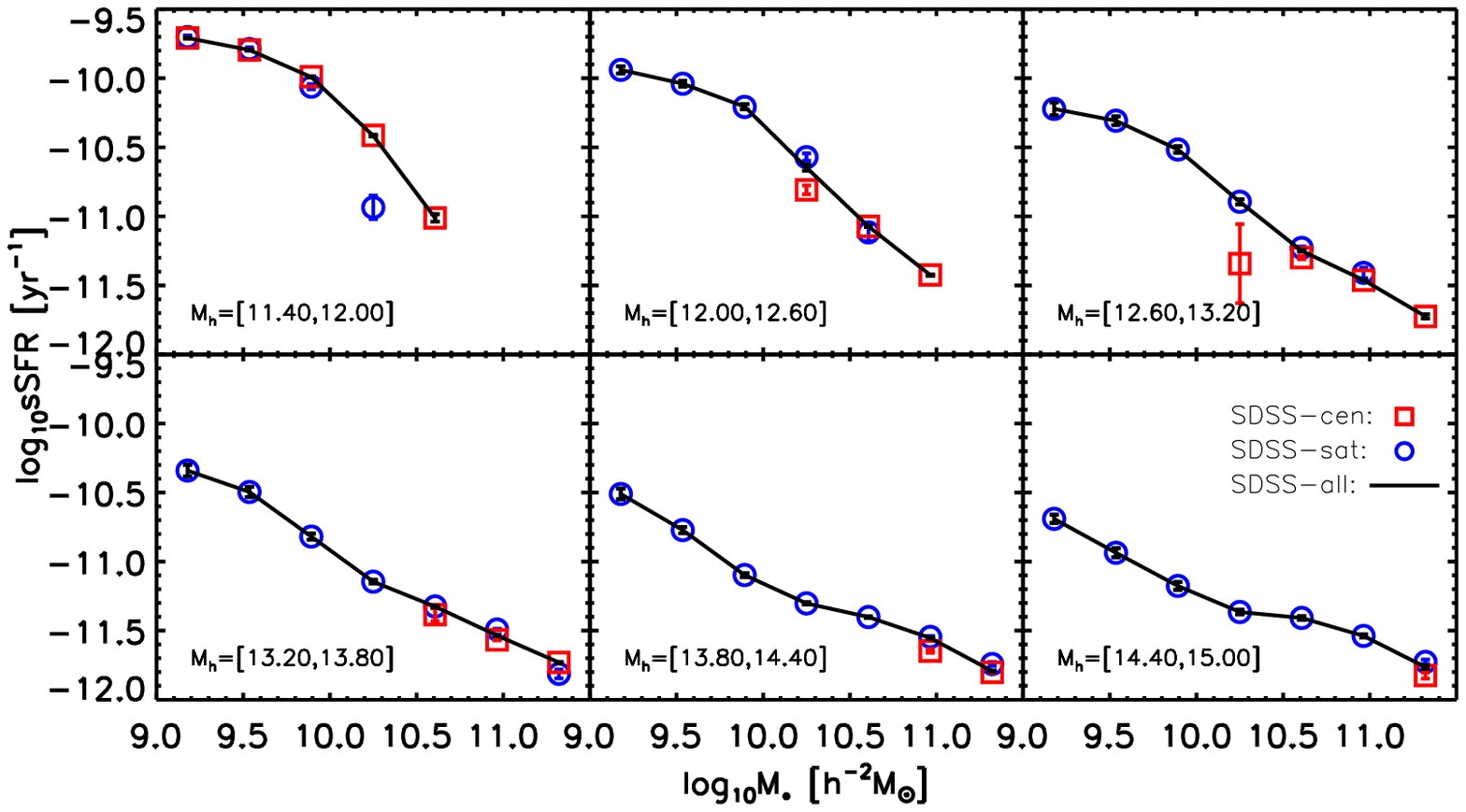,clip=true,width=0.7\textwidth}
    \epsfig{figure=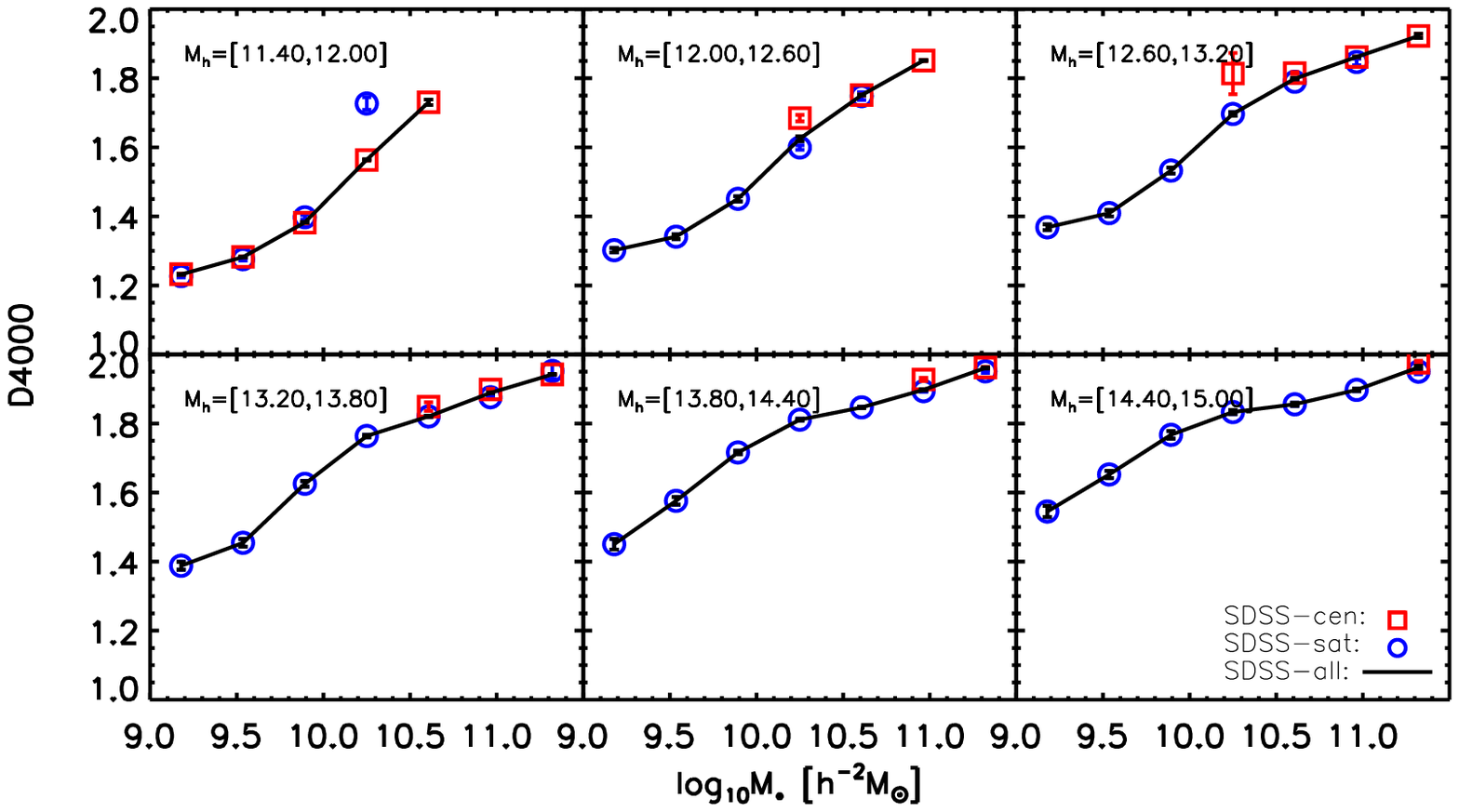,clip=true,width=0.7\textwidth}
    \end{center}
  \caption{The dependence of \dindex\ (top panels) and sSFR (bottom
    panels) on stellar mass for centrals (red squares), satellites
    (blue circles) and all galaxies (black solid line) in a series of
    halo mass bins. Here we present the median value of \dindex\ and
    sSFR for each subsamples. }
  \label{fig:d4000_sSFR_1th}
\end{figure*}

In addition to the quenched fraction, we also present results based on
the specific SFR (sSFR) and the 4000 \AA\ break.  These two quantities
are indicators of the star formation history of a galaxy at different
epochs \citep[e.g.][]{Kauffmann-03,Li-15}.  While the sSFR, defined as
SFR/$M_*$, is sensitive to the strength of very recent (within 50 Myr)
or on-going star formation, the \dindex\ is sensitive to the star
formation of the galaxy over the past 1-2 Gyr.
Figure \ref{fig:d4000_sSFR_1th} plots the median sSFRs and median
\dindex\ of centrals and satellites as functions of stellar mass,
 for different bins in halo mass. Overall, sSFR decreases, and
 \dindex\ increases with increasing stellar mass, in each halo mass
 bin, in agreement with the fact that more massive galaxies tend to
host older stellar populations and are more likely to be
quenched.  More importantly, there is no significant difference in the
stellar mass dependence of sSFR and \dindex\ between centrals and
satellites in a given halo mass bin, indicating that the two
  populations experience, on average, a similar star formation history
  over the past 2 Gyrs.  We have also compared the full distributions
of sSFR and \dindex\ for bins in stellar mass and halo mass, and find
that centrals and satellites have distributions, not just medians,
that are extremely similar.  These results strengthen the notion 
that the star formation and quenching of centrals and
satellites in a halo are governed by the same set of physical
processes.

\subsection{Dependence on $B/T$ and $\sigma_{\rm c}$}
\label{subsec:fq_2th}

\begin{figure*}
  \begin{center}
    \epsfig{figure=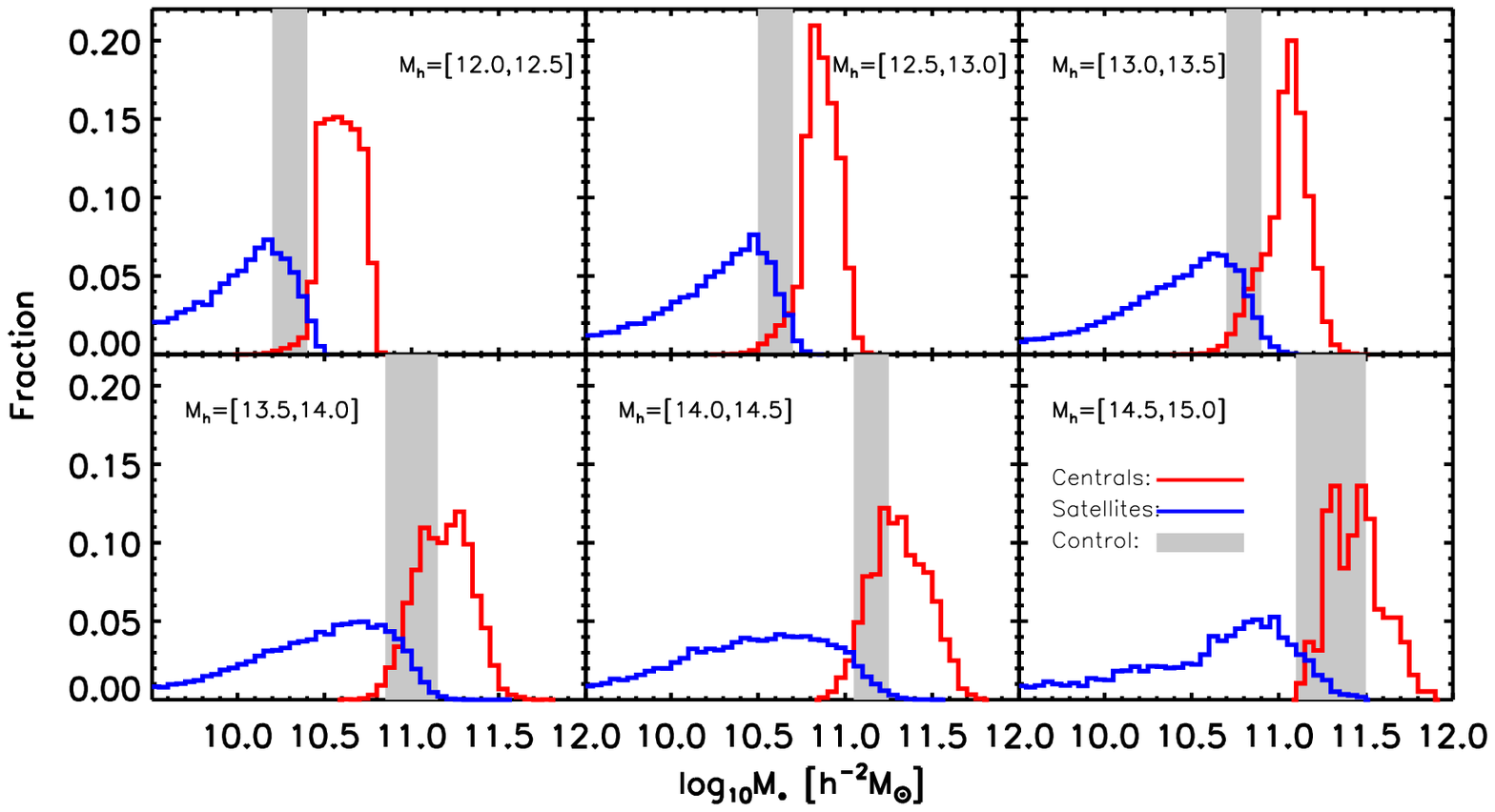,clip=true,width=0.8\textwidth}
  \end{center}
\caption{The normalized distribution of stellar mass for centrals (red
  histograms) and satellites (blue histograms) in six halo mass bins.
  In each panel, the narrow shaded area shows the controlled mass
  range, in which both centrals and satellites have relative high
  abundance. }
 \label{fig:mass_dis}
\end{figure*}

\begin{figure*}
  \begin{center}
   \epsfig{figure=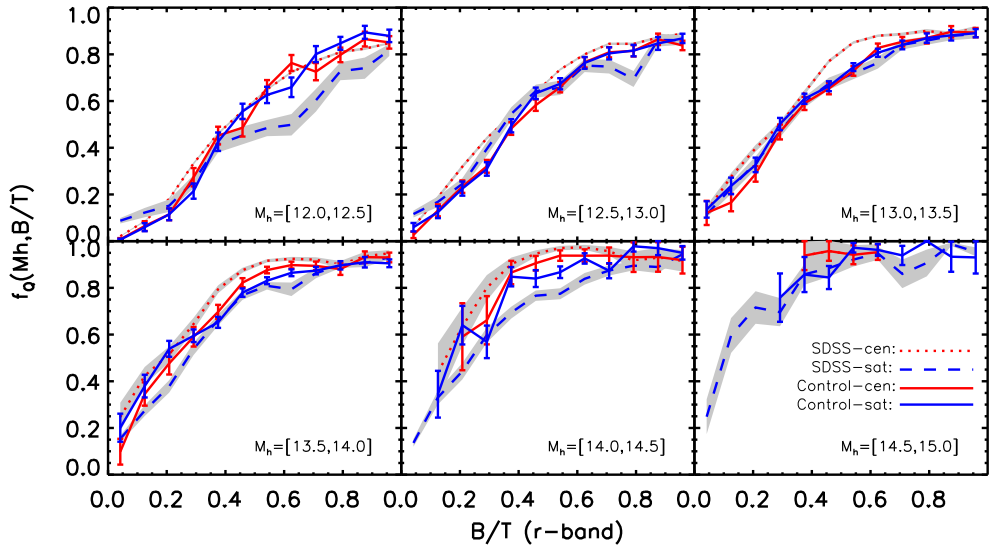,clip=true,width=0.8\textwidth}
  \end{center}
\caption{The quenched fraction as a function of bulge-to-total light
  ratio for centrals and satellites in the six halo mass bins as
  indicated.  In each panel, the red dotted and blue dashed lines with
  shaded regions represent the quenched fraction as a function of
  $B/T$ for all the central and satellite galaxies in the
  corresponding halo mass bin, respectively.  The shaded regions are
  the 1$\sigma$ confidence range.  The red and blue solid lines in
  each panel show the quenched fraction as a function of $B/T$ for
  centrals and satellites in the controlled stellar mass range (as
  indicated in the panel), respectively.}
 \label{fig:quench_bt_2th}
\end{figure*}

\begin{figure*}
  \begin{center}
   \epsfig{figure=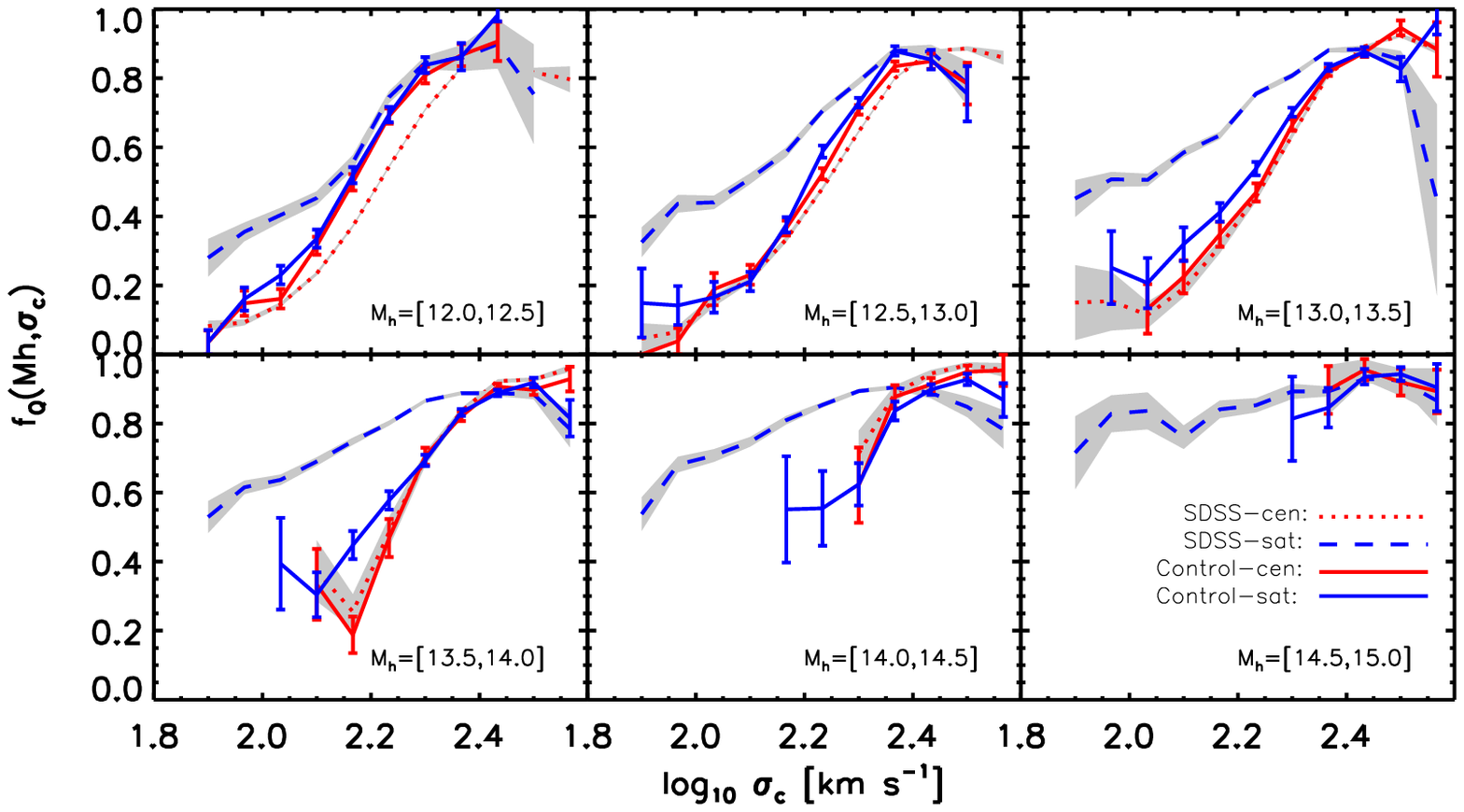,clip=true,width=0.8\textwidth}
  \end{center}
\caption{The quenched fraction as a function of central velocity
  dispersion for centrals and satellites in the six halo mass bins as
  indicated in the panels.  Similar to Figure \ref{fig:quench_bt_2th},
  the red dotted and blue dashed lines with shaded regions in each
  panel represent the quenched fraction as a function of
  $\log_{10}\sigma_c$ for all the central and satellite galaxies in
  the corresponding halo mass bin, respectively. The relations for
  galaxies in the controlled stellar mass bins (as labeled in each
  panel) are shown as the red solid lines for centrals and the blue
  solid lines for satellites.}
 \label{fig:quench_vc_2th}
\end{figure*}

It has been suggested that the structural properties of
  (central) galaxies may be more closely related to the quenched
fraction than stellar mass \citep[e.g.][]{Driver-06, Bell-08,
Cameron-09, Bell-12, Fang-13, Bluck-14, Woo-15, Bluck-16,
Teimoorinia-Bluck-Ellison-16}. Indeed, as shown in Figure \ref{fig:quench_def}, SFR and \dindex\ show strong correlations with $B/T$ and $\sigma_{\rm c}$, indicating that galaxies with more pronounced bulge or higher central velocity dispersion are more likely to be quenched.
%\textbf{Although the link between structural properties and star formation quenching of galaxies have been established for decades \citep[e.g.][]{Fang-13, Bluck-14, Barro-17}, that the existence of dense cores or bulges is the reason or result for star formation quenching is still under debate \citep[e.g.][]{Martig-09, Lilly-Carollo-16}. }
Motivated by this, we
investigate the quenched fraction as a function of the bulge-to-total
light ratio and central velocity dispersion for centrals and
satellites. The goal is to find out whether quenching of star
formation in centrals and satellites depends on the structural
properties in the same way.  Figure \ref{fig:mass_dis} shows the
normalized stellar mass distributions for centrals and
satellites in a series of halo mass bins.  As expected, at given
  halo mass centrals tend to be more massive than
satellites. In order to facilitate a comparison of centrals and
  satellites that are matched in both halo and stellar mass, we
select a narrow stellar mass range in each halo mass bin
(indicated by the shaded region in each panel), where
  centrals and satellites overlap, and which is broad enough such that
  the samples are not too small.  We refer to these samples as the
`controlled' central/satellite samples, in contrast to the parent
samples, to which we refer as `total' central/satellite
samples.

Figure \ref{fig:quench_bt_2th} shows the quenched fraction as a function 
of $B/T$ for centrals and satellites in the six halo mass bins. 
In each panel, the dotted and dashed lines show the results for
  the total central and satellites samples, respectively, while the
  red and blue lines show the results for the corresponding
  controlled samples. In general, the quenched fraction increases
sharply with bulge-to-total light ratio for both centrals and
satellites, which is consistent with the previous findings that a
massive bulge seems to be a necessary condition for quenching of a
central galaxy \citep{Bell-08, Fang-13, Bluck-14, Barro-17}.
Small, but significant differences are apparent between
the `total' central and satellite samples in almost all halo mass bins
%except for the most massive one (see also Figure \ref{fig:quench_vc_2th}).
to the extent that, at fixed bulge-to-total light ratio,
centrals are more likely to be quenched than satellites.  However,
this is almost entirely due to the different stellar mass distributions
of the two populations. Indeed, the differences disappear, as judged
from the error bars, when stellar mass ranges are controlled (red and
blue lines in Figure \ref{fig:quench_bt_2th}). 
This indicates a dependence of the \fq-B/T relation on stellar mass. We have checked this dependence in a series of halo mass bins, and find that the \fq-B/T relation varies a little at low-to-intermediate stellar mass, while significantly changes at the high stellar mass end (\lgmstar$>$11.0). 

Recently, \citet{Teimoorinia-Bluck-Ellison-16} and \cite{Bluck-16}
found that central velocity dispersion is more closely linked to
the quenching of central galaxies than any other property,
including stellar mass, halo mass and bulge mass.  Figure
\ref{fig:quench_vc_2th} shows the quenched fraction as a function of
central velocity dispersion separately for centrals and satellites.
As in Figure \ref{fig:quench_bt_2th}, we present the \fq-$\sigma_{\rm
  c}$ relations for the total (black line), central (dotted line)
and satellite (dashed line) samples, and for the controlled central
versus satellite samples (red versus blue solid lines).  In general,
the quenched fraction increases rapidly with increasing central
velocity dispersion for both centrals and satellites.  For the total
central and satellite samples, significant differences in the
\fq-$\sigma_{\rm c}$ relation are apparent between the two
populations at halo masses below $10^{14.5}h^{-1}{\rm M}_{\odot}$.
This is in good agreement with \cite{Bluck-16}, who found that
satellites as a whole are more frequently quenched than centrals at a
fixed central velocity dispersion.  In addition, the centrals seem to
show a steeper \fq-$\sigma_{\rm c}$ relation than satellites
in almost every halo mass bin, except for the most
massive one.
%Naively one would expect the quenched fraction of
%central galaxies to be more sensitive to central velocity dispersion,
%as it is perhaps related to the central massive black hole that may
%act to quench the star formation in the host galaxy.
However, this is almost entirely due to the different stellar mass
distribution of the two populations, because the \fq-$\sigma_{\rm c}$ relation strongly depends on stellar mass at given halos.
Indeed, when using the controlled samples instead, the differences are almost entirely
eliminated. Hence, we conclude that the data suggests that
  centrals and satellites obey the same \fq-$B/T$ and \fq-$\sigma_{\rm
    c}$ relations, and with similar dependencies on halo mass and
  stellar mass. This furter supports the notion that there is nothing
  special about `being a satellite' versus `being a central'; rather,
  quenching is governed by stellar mass and/or halo mass, with no
  additional dependence on the central vs. satellite nature (but see
  discussion in \S\ref{sec:summary} below.

\subsection{Dependence on halo-centric radius}

\begin{figure*}
  \begin{center}
   \epsfig{figure=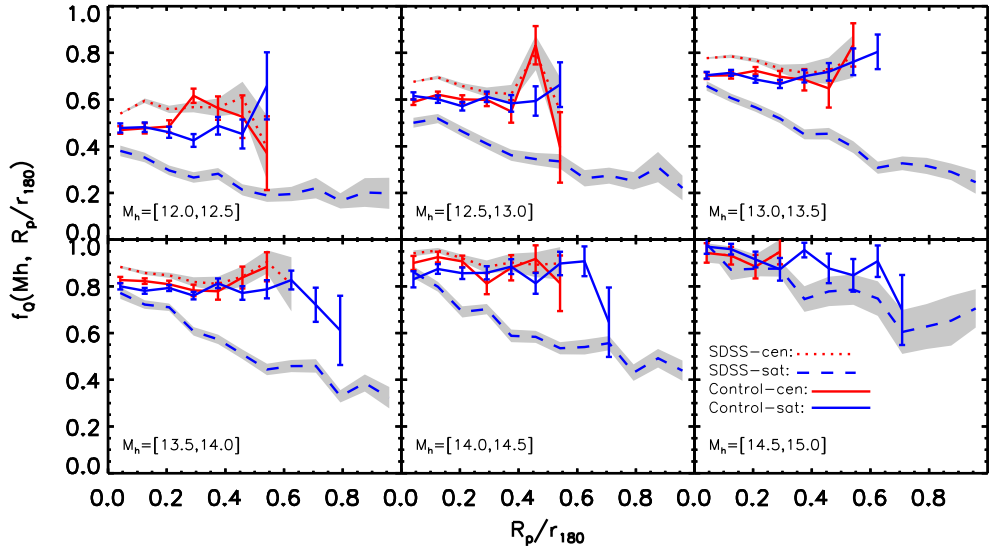,clip=true,width=0.8\textwidth}
  \end{center}
\caption{Quenched fraction as a function of halo-centric radius for
  centrals and satellites in the six halo mass bins.  In each bin,
  relations for all the centrals and satellites are shown in red
  dotted and blue dashed lines, respectively.  The relations obtained
  from the controlled central and satellite samples are shown in red
  and blue solid lines, respectively. The halo mass bin is labeled in
  each panel.}
 \label{fig:quench_pos_2th}
\end{figure*}

\begin{figure*}
  \begin{center}
   \epsfig{figure=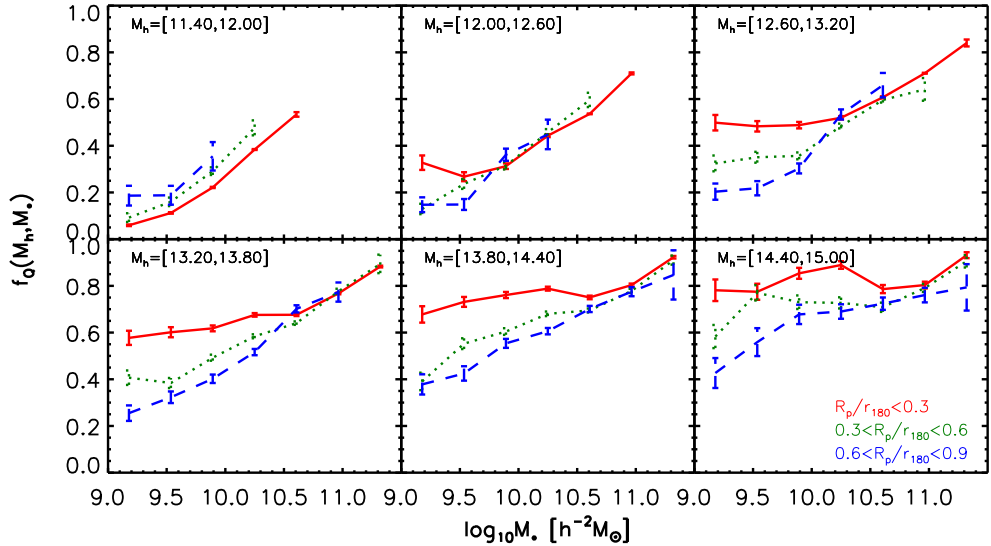,clip=true,width=0.8\textwidth}
  \end{center}
\caption{Quenched fraction as a function of stellar mass in the six
  halo mass bins for galaxies divided into three halo-centric radius
  intervals: $R_{\rm p}/r_{\rm 180}<0.3$ (red solid lines),
  $0.3<R_{\rm p}/r_{\rm 180}<0.6$ (green dotted lines), and
  $0.6<R_{\rm p}/r_{\rm 180}<0.9$ (blue dashed lines).  }
 \label{fig:quench_mstar_1th_radius}
\end{figure*}

The various `satellite-specific' quenching processes discussed
  in the literature (tidal stripping, ram-pressure stripping,
  strangulation, harassment) all are expected to have an efficiency
  that depends on the location of a satellite galaxy within its host
  halo.  Indeed, recent analyses have revealed that the quenched
population of satellite galaxies becomes more dominant
towards the halo center \citep[e.g.][]{Weinmann-06,
  vandenBosch-08b, Wetzel-Tinker-Conroy-12, Woo-13, Kauffmann-13}.
Here we use our data to examine how the quenched fraction depends on
halo-centric distance, and in particular, how centrals and satellites
compare in this regard.

Figure \ref{fig:quench_pos_2th} shows the quenched fraction as a
function of halo-centric radius for both centrals and satellites in
the same six halo mass bins as used in Figures
  \ref{fig:quench_bt_2th} and \ref{fig:quench_vc_2th}. The
halo-centric radius is defined as the projected distance of a
galaxy to the (luminosity-weighted) group center
\citep{Yang-07}, scaled by the halo virial radius. The latter
corresponds to the radius within which the dark matter halo has an
overdensity of 180 (see Equation 2).\footnote{Since the group center
  is defined as the luminosity weighted position of all group members,
  the central galaxy is not necessarily located at the center of the
  group if it has one or more satellites.}  As in Figures
\ref{fig:quench_bt_2th} and \ref{fig:quench_vc_2th}, we display the
relation for the total central and satellite samples in dotted and
dashed lines, respectively, and for the controlled centrals and
satellites samples in red and blue lines, respectively. For the total
satellite sample, the quenched fraction depends strongly on the
halo-centric radius, with the quenched fraction decreasing with
$R_{\rm p}/r_{\rm 180}$. These results indicate that galaxies in the
inner region of halos may be more affected by environmental processes
than galaxies in the outer regions, as expected if quenching arises
from, for example, tidal interactions or ram-pressure stripping. 
Since B/T is found to be a good predictor of star formation quenching, we have examined the B/T as a function of halo-centric radius in a series of stellar mass bins at given halo mass. We find that the shape of \fq-$R_{\rm p}/r_{\rm 180}$ relation resembles the shape of B/T-$R_{\rm p}/r_{\rm 180}$ relation as a whole, suggesting that the \fq\ dependence on halo-centric radius could be explained by the B/T dependence of halo-centric radius. However, that the existence of massive bulge is the driving factor or the by-product of star formation quenching is still under debate \citep[e.g.][]{Martig-09,Lilly-Carollo-16, Wang-18b}. 

For the total central sample, on the other hand, the $R_{\rm p}$
dependence is rather weak. However, for the controlled samples,
these differences between centrals and satellites are almost
  entirely eliminated, and the quenched fractions for both the
central and satellite populations show no significant $R_{\rm
  p}$-dependence. Note that satellites in the controlled sample are
located in the high mass tail of the distribution of the total sample
(see Figure \ref{fig:mass_dis}). It is conceivable that more massive
satellites are less affected by environmental effects than the less
massive ones, which may be the reason why the \fq-$R_{\rm p}/r_{\rm
  180}$ relation seen for the controlled satellite sample is
flat.  This result is also consistent with \cite{Wang-18} who found
that the dependence of the quenched fraction on halo-centric radius
appears only for galaxies with masses much lower than that of the
central galaxies in their corresponding host halos.
%We note that there are a small fraction of central galaxies beyond
%$R_{\rm p}/r_{\rm 180}\sim0.4$, even at the lowest halo mass bin,
%while these centrals are rare cases according to the large error bars
%shown in Figure \ref{fig:quench_pos_2th}.

Since the quenching fractions of centrals and satellites of similar
stellar mass do not show significant differences in their dependencies
on $B/T$ or halo-centric radius, in halos of similar mass, we
can look at the \fq-\mstar\ relation without separating centrals and
satellites, but for galaxies divided according to their halo-centric
distances.  Figure \ref{fig:quench_mstar_1th_radius} shows the
quenched fraction versus stellar mass for three intervals of $R_{\rm
  p}/r_{\rm 180}$.  As one can see, the quenched fraction only depends
significantly on halo-centric distance at the low stellar mass end,
with galaxies located closer to the group center being more likely to
be quenched. This result is in good agreement with previous studies
\citep[e.g.][]{Weinmann-06, Wetzel-Tinker-Conroy-12, Woo-15, Wang-18}.
Note that, for a given halo mass bin, there appears to be a
  stellar mass threshold, above which the quenched fraction becomes
independent of halo-centric distance. This stellar mass threshold
increases with increasing halo mass, from $\sim
10^{9.7}h^{-2}\Msun$ to $\sim10^{10.5}h^{-2}\Msun$, as the halo mass
increases from $10^{12}h^{-1}\Msun$ to
$10^{15}h^{-1}\Msun$. This agrees with the fact that the
  quenched fractions of centrals and massive satellites do not show
  significant dependence on halo-centric distance (cf. Figure
  \ref{fig:quench_pos_2th}). It also suggests that, as long as the
stellar mass of a galaxy is sufficiently large, the probability for it
to be quenched does not depend on its location in the host halo, no
matter whether it is a central or a satellite.  
%\comment{I suggest removing the sentences below (`Finally, for....higher mass groups'), as they don't really add anything, other than confusion.}
%Finally, for the lowest halo mass bin (\lgmhalo$<12$), the $f_{\rm Q}$- $M_*$ relations for different $R_p$ bins do not seem to converge at the high $M_*$ end. However, for groups of such low mass, galaxies located in the inner part appear to have a lower quenched fraction than in the outer parts over the entire stellar mass range covered, a trend in contrast with that seen for galaxies in higher mass groups.

\section{Results of AGN activities}
\label{sec:AGN}

\begin{figure*}
  \begin{center}
    \epsfig{figure=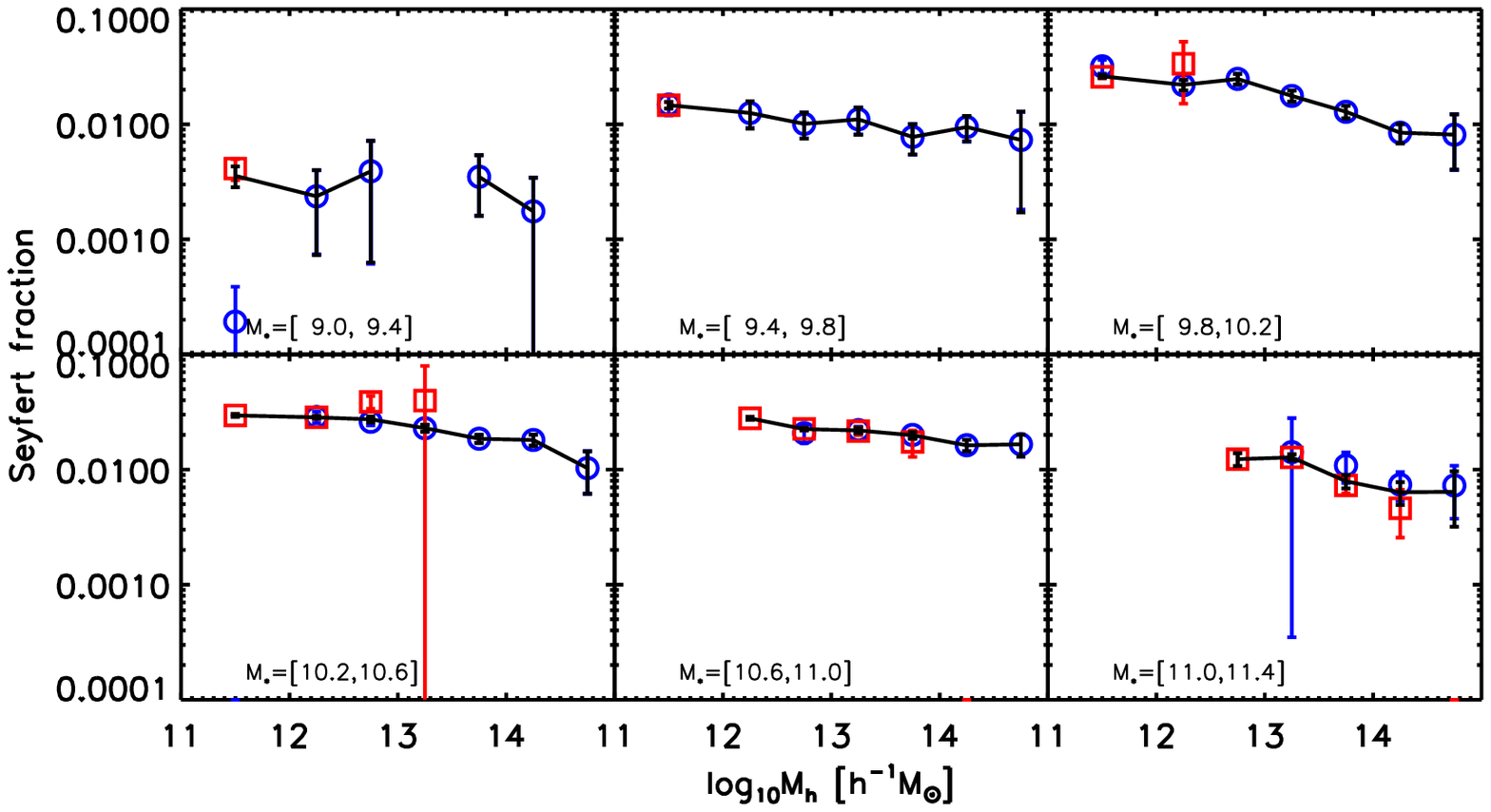,clip=true,width=0.7\textwidth}
    \epsfig{figure=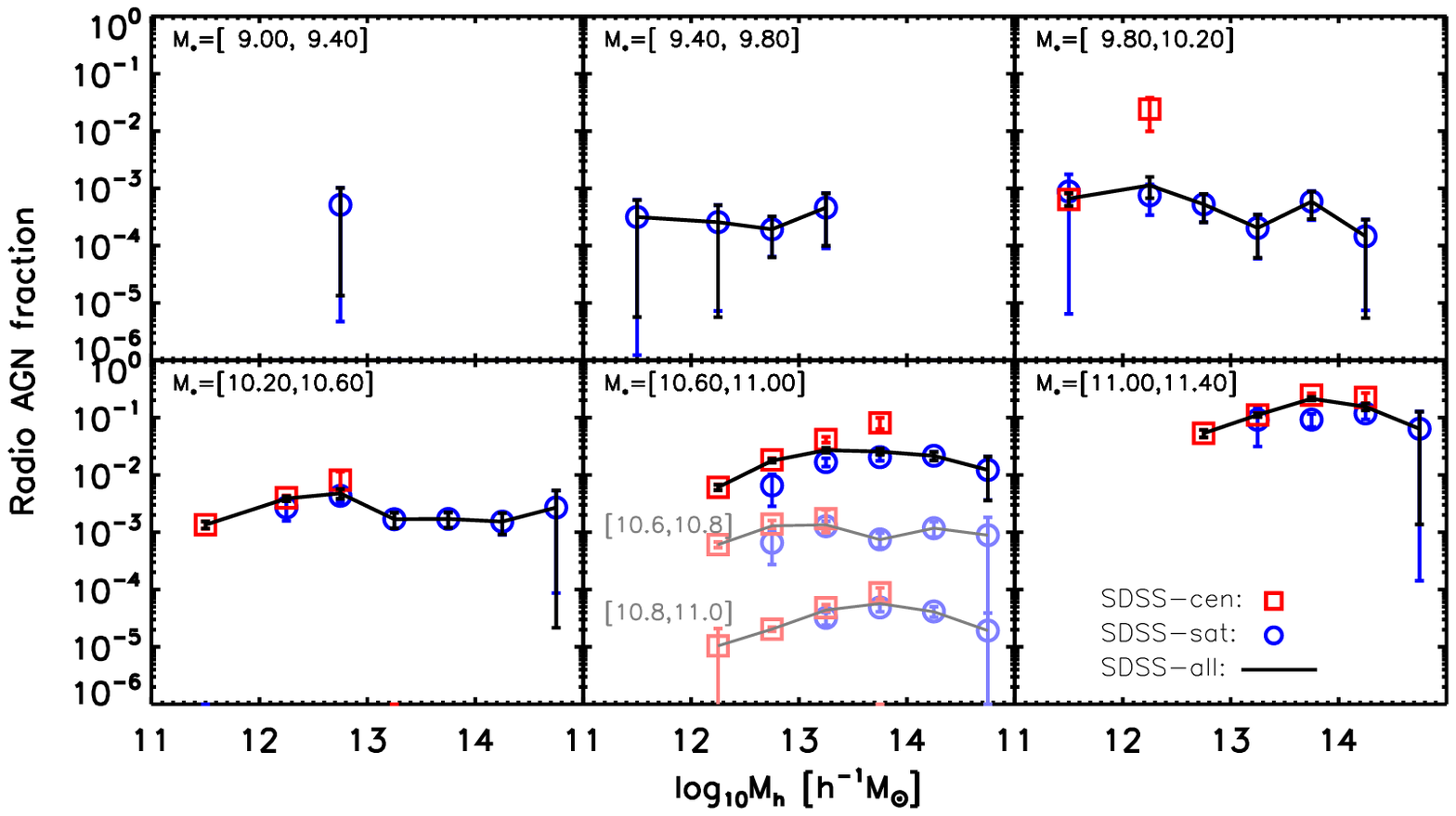,clip=true,width=0.7\textwidth}
    \end{center}
  \caption{The dependence of the optical-selected (top panels) and the
    radio-loud (bottom panels) AGN fractions on halo mass for centrals
    (red squares), satellites (blue circles) and all galaxies (black
    lines) in a series of stellar mass bins, as indicated. The errors
    are estimated by bootstrap method.  In the bottom panel of
    $10.6<$\lgmstar$<11.0$, we further divide galaxies into two
    narrower stellar mass bins, $[10.6,10.8]$ and $[10.8,11.0]$, and
    present the radio-loud AGN fractions of the two subsamples,
    shifted by 1 dex and 3 dex, respectively, .}
  \label{fig:agn_frac_mh}
\end{figure*}

\begin{figure*}
  \begin{center}
    \epsfig{figure=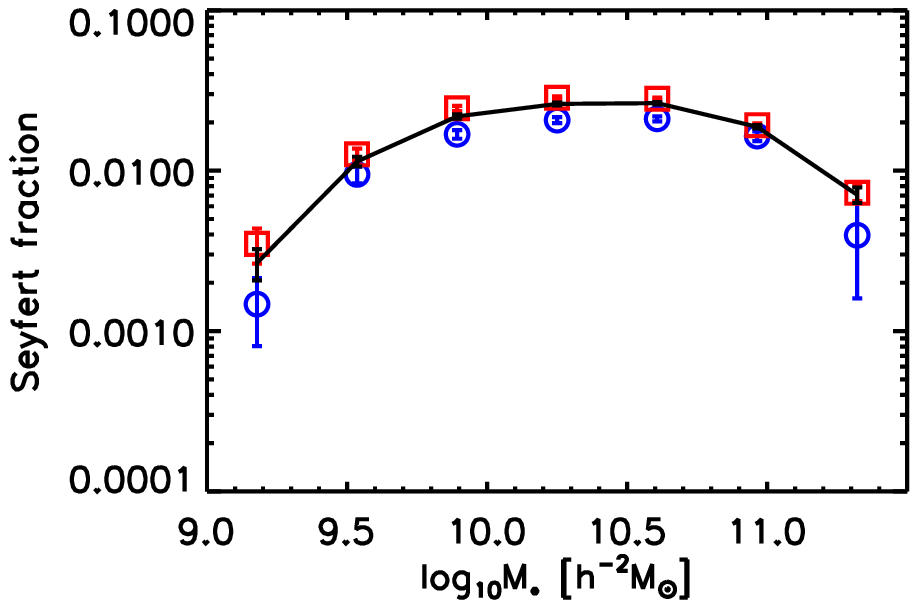,clip=true,width=0.45\textwidth}
    \epsfig{figure=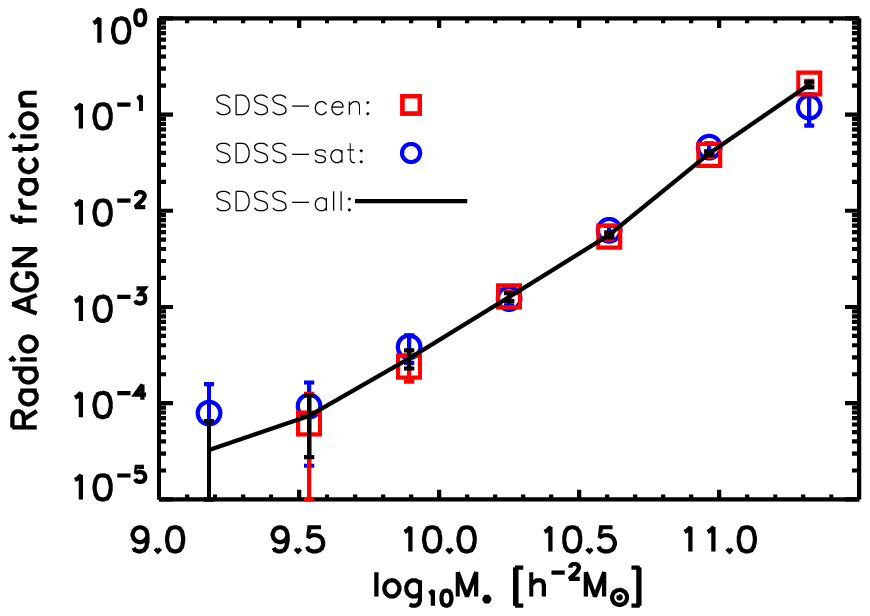,clip=true,width=0.45\textwidth}
    \end{center}
  \caption{Optical-selected (left panel) and radio-loud (right panel)
    AGN fraction as a function of stellar mass for centrals (red
    squares), satellites (blue circles) and all galaxies (black
    lines).}
  \label{fig:agn_frac_mstar}
\end{figure*}

AGN feedback has been suggested as an important internal quenching
process in many galaxy formation models, although its details and
efficiency are still unclear and under debate. To shed light on the
problem, it is interesting to check whether centrals and satellites
have different AGN properties.  Here we investigate the populations of
optical-selected and radio-loud AGNs in both centrals and satellites.

\subsection{The Seyfert fraction}

The top panels of Figure \ref{fig:agn_frac_mh} shows the fraction of
optical-selected AGNs (Seyfert fraction) as a function of halo
mass for centrals and satellites in different stellar mass bins. As
one can see, the fraction of Seyfert galaxies decreases with
increasing halo mass, albeit slowly, in almost all the stellar mass
bins. In addition, centrals and satellites of similar stellar masses
do not show any significant difference in the Seyfert fraction at
fixed halo mass. Since the Seyfert fraction depends only weakly on
halo mass for both centrals and satellites, Figure
  \ref{fig:agn_frac_mstar} presents the results as a function of
stellar mass, without binning in halo mass. As one can see, the
fraction of optical-selected AGN peaks around a stellar mass of
  $\sim 10^{10} - 3 \times 10^{10} h^{-2} \Msun$, in good agreement
  with the results of \cite{Pasquali-09}.  
%This is likely due to the combined effect of increase in black halo
%mass and decrease in cold gas fueling as the stellar mass increases.
In addition, at a given stellar mass, centrals appear to have a
slightly higher AGN fraction than satellites. This is caused by the
weak anti-correlation between the AGN fraction with halo mass seen in
Figure \ref{fig:agn_frac_mh}, and indeed the difference between
centrals and satellites are eliminated when comparisons are made
within narrow halo mass bins.  The fraction of Seyfert galaxies among
centrals and satellites is about 2\%-3\% in the intermediate stellar
mass range, between $10^{10.0}$ and $10^{10.5}h^{-2}\Msun$, in broad
agreement with the percentage, 3\%-4\%, found in \cite{Pasquali-09}
using a different definition of optical AGNs.

We also examine the dependence of Seyfert fraction on the halo-centric
distance without distinguishing centrals from satellites, and the
result is shown in the left panel of Figure \ref{fig:agn_radius}.
Remarkably, there is only a very weak dependence of Seyfert
fraction on halo-centric distance; galaxies located closer towards
the centers of their halos have a slightly enhanced probability of
being Seyfert galaxies. However, the slightly higher Seyfert
fraction seen in the inner bin is consistent with the slightly
higher Seyfert fraction among central galaxies shown in the left
panel of Figure \ref{fig:agn_frac_mstar}, combined with the fact
that the innermost radial bin has a much higher central fraction
than the other bins.  Hence, this slight enhancement ultimately owes
to the weak anti-correlation between the Seyfert fraction and halo
mass seen in Figure \ref{fig:agn_frac_mh}. No significant
difference is found for the other two bins, $0.3<R_{\rm p}/r_{\rm
  180}<0.6$ and $0.6<R_{\rm p}/r_{\rm 180}<0.9$.  We thus conclude
that the Seyfert fraction of galaxies does not depend significantly on
halo-centric distance in halos of similar masses, and that there
  is no discernable difference in the Seyfert fraction of centrals and
  satellites.

\subsection{Radio-loud AGN fraction}

The bottom panels of Figure \ref{fig:agn_frac_mh} show the radio AGN
fraction as a function of halo mass for centrals and satellites in a
series of stellar mass bins. In all stellar mass bins shown,
the radio-loud AGN fraction of centrals and satellites only
  depends very weakly on halo mass, if at all. This is in stark
  contrast to the strong dependence on stellar mass, which is evident
  from a comparison of the different panels, and is depicted more
  clearly in the right panel of Figure \ref{fig:agn_frac_mstar} and
  discussed below.  As for the Seyfert and quenched fractions, the
  radio loud fractions of centrals and satellites (matched in both
  stellar and halo mass) are virtually indistinguishable. An exception
  is the stellar mass bin of $10.6<$\lgmstar$<11.0$, where centrals
  appear to have a slightly enhanced radio loud fraction. However,
the large difference seen in this stellar mass bin is largely due to
the different stellar mass distribution between the two populations.
In this bin, the number of satellites shows a rapid decrease, while
that of centrals shows a rapid increase, with increasing stellar mass,
so that difference in the stellar mass distribution is significant
between the two populations.  Indeed, when galaxies are further
divided into two narrower stellar mass bins, [10.6,10.8] and
[10.8,11.0], the difference in radio-loud AGN fraction between
centrals and satellites in each of the two bins disappears.

\begin{figure*}
  \begin{center}
    \epsfig{figure=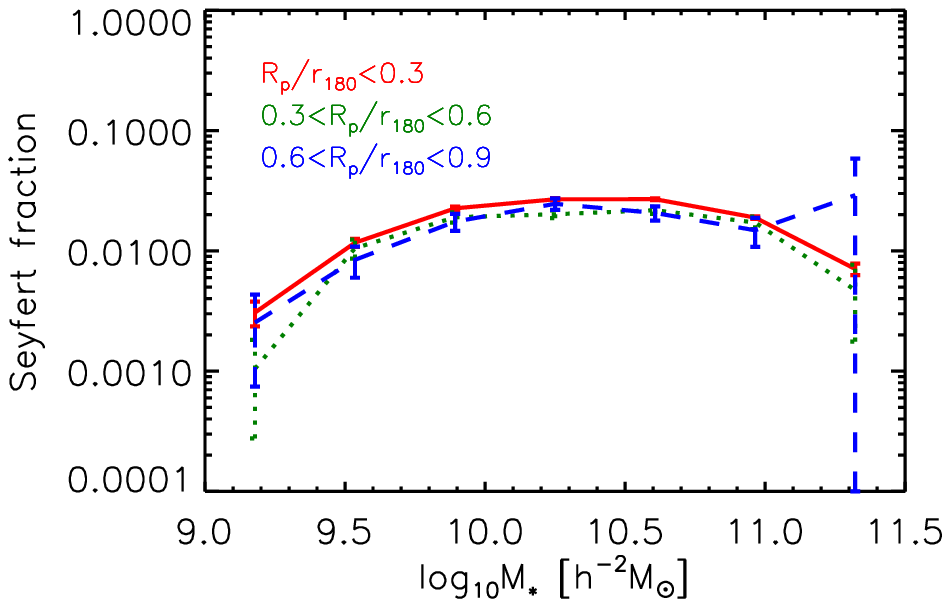,clip=true,width=0.45\textwidth}
    \epsfig{figure=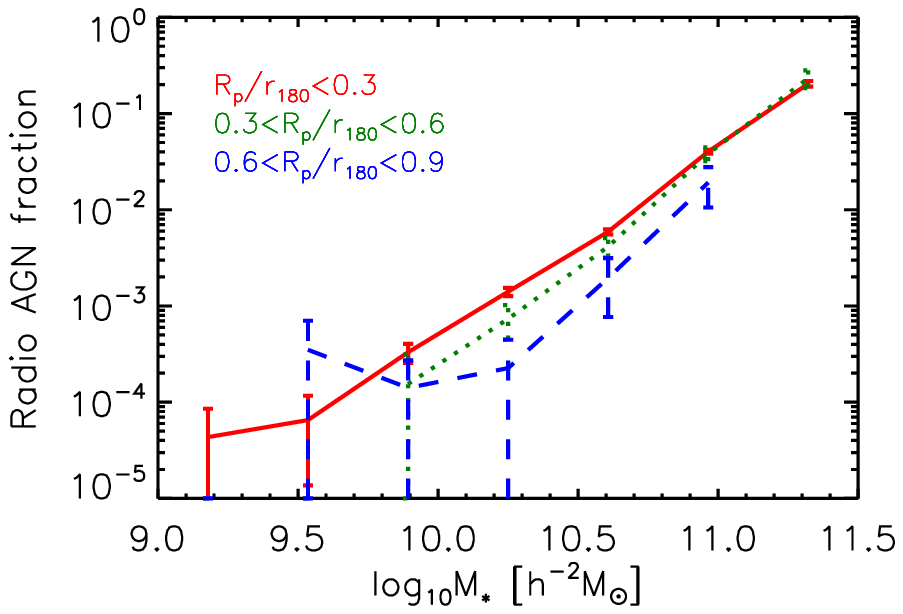,clip=true,width=0.45\textwidth}
    \end{center}
  \caption{Optical-selected (left panel) and radio-loud (right panel)
    AGN fraction as a function of stellar mass by dividing galaxies
    into three halo-centric radius intervals: $R_{\rm p}/r_{\rm
      180}<0.3$ (red solid lines), $0.3<R_{\rm p}/r_{\rm 180}<0.6$
    (green dotted lines) and $0.6<R_{\rm p}/r_{\rm 180}<0.9$ (blue
    dashed lines). }
  \label{fig:agn_radius}
\end{figure*}

Since the fraction of radio-loud AGN exhibits no significant
dependence on halo mass at fixed stellar mass, we show, in the right
panel of Figure \ref{fig:agn_frac_mstar}, the fraction of radio-loud
AGN as a function of stellar mass without separating galaxies into
different halo mass bins. For both centrals and satellites, the
radio-loud AGN fraction is strongly correlated with stellar mass
\citep[see also][]{Best-05}.  The radio-loud fraction increases by
more than three orders of magnitude over the stellar mass range
covered by our sample. More importantly, there is no significant
difference between centrals and satellites over the entire stellar
mass range. This is in good agreement with \cite{Pasquali-09},
  who also used the \cite{Yang-07} group catalog, but appears to be
in conflict with the results of \cite{Best-07}, who found that
brightest cluster galaxies (BCGs) are more likely to host a
radio-loud AGN than other galaxies of the same stellar mass. This
inconsistency may be due to (1) a difference between our
  `central' galaxies and the BCGs identified by \cite{Best-07}, and
  (2) the lack of a rigorous control sample in the analysis of
  \cite{Best-07}.  In fact, the BCGs used by \cite{Best-07} are
  defined as the galaxies closest to the deepest point of the
  gravitational potential well of the cluster \citep[see][for
    details]{vonderLinden-07}.  However, the central galaxies used
  here, are simply defined to be the brightest group members, which do
  not necessarily reside at the bottoms of their gravitational
  potential wells \citep[][]{vandenBosch-05, vonderLinden-07,
    Skibba-11}. As shown in the right panel of Figure
\ref{fig:agn_radius}, the radio-loud fraction does depend
weakly, but significantly, on halo-centric distance, in the
sense that galaxies in the inner regions of groups/clusters are more
likely to host a radio-loud AGN than those of similar stellar
mass in the outer regions.  The dependence appears to be stronger in
the intermediate stellar mass range and becomes insignificant at the
massive end. Although the statistics are definitely poor, this may
explain the discrepancy between our results and those of \cite{Best-07},
if a significant fraction of the centrals they identified are not the
most massive cluster galaxies.

Based on the results presented above, we conclude that the likelihood
for a central galaxy (defined as the most massive group member)
to be an optical or radio-loud AGN is similar to that of a satellite
with similar stellar masses. This suggests that both the central
  engine (the supermassive black hole) and the fuel-supply are
similar for centrals and satellites that are matched in both
  stellar and halo mass.

\section{Summary and discussion}
\label{sec:summary}

It is well established that central and satellite galaxies of
  the same stellar mass have substantially different properties. In
  particular, satellites are more likely quenched. This is often
  interpreted as evidence for `satellite-specific' quenching
  mechanisms, such as strangulation, ram-pressure stripping, or galaxy
  harassment. However, at fixed stellar mass, satellite galaxies
  typically reside in much more massive halos than central
  galaxies. Hence, an alternative explanation for the difference could
  be that there is nothing special about being a satellite versus
  being a central, but there is a strong dependence of quenching on
  galaxy environment (halo mass in particular). In this paper we have
  carried out a detailed comparison between central and satellite
  galaxies with regard to their starformation and AGN activity. In
  particular, we used the SDSS galaxy
  group catalog of \cite{Yang-07} to examine the quenched fractions
  and AGN fractions of centrals and satellites as functions of galaxy
  stellar mass, host halo mass, galaxy structural properties and
  halo-centric distances. In order to break the degeneracy between the
  two `alternative interpretations' mentioned above, and to assess to
  what extent `being a central' versus `being a satellite' impacts the
  star formation and/or AGN activity, we compared central and
  satellite samples that are matched in both stellar mass and halo
  mass. Our main results are:

\begin{itemize}

\item Centrals and satellites show similar \fq-$M_{\rm h}$ and
  \fq-$M_\ast$ relations.  This strongly suggests, but does not prove,
  that centrals and satellites of similar stellar mass experience
  similar quenching processes. Moreover, the median sSFR and
  \dindex\ of the two populations are similar at given stellar mass
  and halo mass, suggesting similar star formation histories.

\item Tight and strong correlations of the quenched fraction with the
  bulge-to-total light ratio and central velocity dispersion are found
  for both centrals and satellites.  When both halo mass and stellar
  mass are controlled, centrals and satellites follow identical
  \fq-$B/T$ and \fq-$\sigma_{\rm c}$ relations.

\item The quenched fraction of centrals exhibits weak or no
  correlation with the halo-centric distance ($R_p/r_{180}$), while
  satellites, in a given bin of halo mass, show a decreasing
  trend in the quenched fraction from the group center
  outward. However, when controlling for both stellar and halo
    mass, the quenched fraction is once again similar for both
  centrals and satellites.

\item Satellite galaxies with stellar masses that are comparable
  to that of their centrals, have quenched fractions that are not
  correlated with halo-centric radius. Less massive satellites,
  however, reveal a clear trend whereby the quenched fraction
  increases with decreasing halo-centric distance.

\item The fraction of central and satellite galaxies that host
  an optical or radio-loud AGN depends strongly on stellar mass, but
  only very weakly on halo mass, in excellent agreement with the previous findings. In addition, when controlling for
  both stellar and halo mass, centrals and satellites have
  optical/radio-loud AGN fractions that are indistinguishable. All
  these results suggest that triggering AGN activity has little to no
  dependence on halo mass, or on being a central versus a satellite.
  
%At a given stellar mass, the presence of optical/radio-loud AGNs in a
%galaxy shows weak correlations with halo mass. Centrals and satellites
%have similar factions of optical/radio-loud AGNs at given stellar
%mass, suggesting that the effects of AGN feedback and fueling are
%similar for the two populations.

\end{itemize}

To summarize, we confirm the finding of \cite{Wang-18} that, in
halos of a given halo mass, the quenched fraction for central galaxies
is similar to that for satellites of the same stellar mass. In
addition, we have demonstrated that the dependence of the quenched
fraction on bulge-to-total light ratio, $B/T$, central velocity
dispersion, $\sigma_{\rm c}$, and halo-centric distance, $r_{\rm p}$
for satellites are indistinguishable from the same relations for
centrals, once galaxy stellar mass and host halo mass are controlled
for in the comparison. These results suggest that centrals and
satellites are indistinguishable in their star formation quenching,
namely
\begin{equation}
f_{\rm Q,cen}(M_*, M_{\rm h}, r_{\rm p}, B/T, \sigma_{\rm c})
=
f_{\rm Q,sat}(M_*, M_{\rm h},  r_{\rm p}, B/T, \sigma_{\rm c})\,.
\end{equation}

Earlier studies by \cite{Bell-08, Cheung-12, Fang-13, Bluck-14,
  Bluck-16} show that the quenched fraction of central galaxies
depends strongly on the central velocity dispersion and bulge-to-total
light ratio.  In particular, the central velocity dispersion is found to be the most relevant quantity for quenching among a number of other quantities considered \citep{Teimoorinia-Bluck-Ellison-16}.  
Since $B/T$ and $\sigma_{\rm c}$ are both internal properties of galaxies, their strong relations to quenching may indicate that quenching is predominantly be driven by internal processes. We find that the \fq-$\sigma_{\rm c}$ and \fq-$B/T$ relations for centrals and satellites are similar once their stellar masses and host halo masses are properly controlled.  We also find that the optical/radio-loud AGN fractions in centrals and satellites are also similar. The two results together suggest that the internal quenching processes may operate in a similar way in both centrals and satellites. %These results have important implications for AGN feedback models adopted in current galaxy formation models. For example, the radio AGN feedback, which is usually thought to play an important role at low redshift and in massive galaxy clusters, should work on both centrals and satellites with almost the same efficiency.

The dependence on the halo-centric distance shown in Figure
\ref{fig:quench_pos_2th} is usually considered as an evidence for
environmental effects in galaxy quenching. When a galaxy falls into a
galaxy group or a galaxy cluster, it is expected to suffer from a series of environmental effects, such as strangulation, tidal stripping, ram-pressure stripping, and merging with companions \citep[e.g.][]{Gunn-Gott-72,Moore-96,Cox-06,Read-06,Boselli-Gavazzi-06,  Weinmann-09}.  These processes, which are thought to operate only on satellites, are expected to be more effective in the central parts of galaxy groups than in the outer regions. However, as pointed out by \cite{Wang-18}, if these processes (or part of them) are indeed the dominant processes quenching star formation in satellites, it would be difficult to understand the similarity in the quenched fraction between centrals and satellites. Our results show that the dependence of the quenched fraction on halo-centric distance is very weak for massive galaxies, regardless whether they are classified as centrals or satellites. 
%As shown in \cite{Wang-18}, the quenched fraction is independent of halo-centric distance for galaxies with masses larger than about one fifth of the mass of the central galaxy in their host halo. Such a dependence on galaxy mass is expected in some satellite-specific processes, in which the effects are expected to be stronger for lower-mass galaxies \citep[e.g.][]{Gunn-Gott-72,Moore-96,Cox-06,Read-06,Boselli-Gavazzi-06,   Weinmann-09}.  Our results, therefore, may provide constraints on these processes.

The dependence on halo mass can be produced by various processes,
including both internal and environmental processes. If the accretion
rate of radio AGNs is positively correlated with the hot gas mass in
halo, as is assumed in some semi-analytic models
\citep[e.g.][]{Croton-06, Guo-11, Henriques-15}, the efficiency of
radio AGN feedback is expected to increase with host halo
mass. Ram-pressure stripping can also produce a halo-mass dependence,
since it is expected to be more important in higher mass halos that
contain more hot gas. In addition, shock heating, which heats cold gas
through accretion shocks and subsequently reduces gas supply for
further star formation, is another mechanism to quench star formation
\citep[e.g.][]{Rees-Ostriker-77, Dekel-Birnboim-06}.  This quenching
process is also expected to be more important in higher mass halos
where gas can be heated to higher temperatures.  More recently,
\cite{Gabor-Dave-15} found that a recipe, in which quenching occurs in
regions dominated by hot gas ($>10^{5.4}$K) in the hydro-dynamical
simulations, can roughly reproduce the trends of the quenched fraction
with halo mass, stellar mass and halo-centric
distances. Interestingly, their simulations also show that the hot gas
has the same quenching effect on centrals and satellites, which is
consistent with our results.

We note the readers that there are also alternative explanations for our results. 
%our results do not imply that satellite-specific quenching processes do not operate. 
If a central galaxy in a halo of mass $M_1$ falls into, and thus becomes a satellite of, a much bigger halo of mass $M_2 \gg M_1$, it is likely to experience a variety of processes that may cause it to quench. And some off these processes may well be `satellite-specific' in that they only operate on satellite galaxies, with ram-pressure stripping being a good example. However, if the galaxy falls into a halo that is only slightly more massive (i.e., $M_2 \sim M_1$), our results suggest that this accretion (transition from being a central to being a satellite) has little to no influence on the galaxy's star formation and/or AGN activity. This is not too surprising. After all, when two halos of comparable mass merge, the dynamical friction time is relatively short, and one expects the two systems to quickly merge. Hence, if we catch the system prior to coalescence of the two galaxies, it means we probably observe the system fairly shortly after the satellite was accreted, and there may simply not have been enough time for (satellite-specific) quenching processes to operate. Indeed, \cite{Wetzel-13} have advocated a fairly long (2-4 Gyr) `delay' time between accretion and the onset of (rapid) quenching.  However, it is unclear whether this mechanism can establish the
similar dependence of quenching efficiency on halo mass for centrals and satellites \citep[See e.g.][]{Wang-18}.

Another potential explanation for our results relates to group finder errors. As pointed out in \cite{Campbell-15}, group finders are not perfect, and introduce a variety of systematic errors due to the combined effect of errors in group membership determination, central/satellite designation, and halo mass assignments. One of the main tendencies of these errors is to reduce the real difference between centrals and satellites, and to make them appear more similar than they really are. 
%Because of these potential shortcomings we refrain from a more detailed discussion regarding the implications of our results for galaxy quenching.  
In the second paper of this series (Wang et al., 2018, in preparation), we compare the results obtained here with predictions of the L-GALAXIES semi-analytic galaxy formation model developed by \cite{Henriques-15}, and the state-of-the-art hydro-dynamic EAGLE (Evolution and Assembly of GaLaxies and their Environments simulation presented in \cite{Schaye-15}. In particular, in order to facilitate a fair and meaningful comparison, we run the same halo-based group finder of \cite{Yang-05} over mock data sets constructed from these models and examine (i) to what extent the group finders wash away potential differences between true centrals and true satellites, and (ii) to what extent L-GALAXIES and EAGLE are able to reproduce the SDSS data.

%The discussion given above serves only as a qualitative analysis of the implications of our results for potentially important quenching processes. In the second paper of this series, we compare the observational results obtained here with the predictions of a semi-analytic galaxy formation model, L-GALAXIES, developed by \cite{Henriques-15}, and a state-of-the-art hydro-dynamic simulation, EAGLE (Evolution and Assembly of GaLaxies and their Environments) presented in \cite{Schaye-15}.  There we will examine in detail how our results can provide insights into star formation and quenching processes in galaxies.

\acknowledgments

This work is supported by the National Natural Science Foundation of
China (NSFC, Nos.  11522324,11733004, 11421303, 11433005, and
11320101002), the National Basic Research Program of China (973
Program)(2015CB857002, 2015CB857004), and the Fundamental Research
Funds for the Central Universities.  EW acknowledges the support from
the Youth Innovation Fund by University of Science and Technology of
China (No. WK2030220019) and China Postdoctoral Science Foundation funded project
(No. BH2030000040). SHC is supported by the Fund for Fostering Talents
in Basic Science of the National Natural Science Foundation of China
NO.J1310021. HM also acknowledges the support from NSF AST-1517528 and
NSFC-11673015. FvdB is supported by the Klaus Tschira Foundation, by
the US National Science Foundation through grant AST 1516962, and by
the National Aeronautics and Space Administration through NASA-ATP
grant 80NSSC18K0524.

\bibliography{rewritebib.bib}

\clearpage

%%%%%%%%%%%%The End%%%%%%%%%%%%%%%%%%%%%%%%%%%%%%%%%%%%%%%%%
\label{lastpage}
\end{document}